\documentclass{mn2e}
\usepackage{graphicx}
\input{epsf}

\newif\ifAMStwofonts
\AMStwofontstrue

\newcommand{\go}{\mathrel{\raise.3ex\hbox{$>$}\mkern-14mu
             \lower0.6ex\hbox{$\sim$}}}
\newcommand{\lo}{\mathrel{\raise.3ex\hbox{$<$}\mkern-14mu
             \lower0.6ex\hbox{$\sim$}}}
\newcommand{\be}{\begin{equation}}
\newcommand{\ee}{\end{equation}}
\newcommand{\ba}{\begin{eqnarray}}
\newcommand{\ea}{\end{eqnarray}}
\newcommand{\lp}{\left(}
\newcommand{\rp}{\right)}
\newcommand{\lb}{\left[}
\newcommand{\rb}{\right]}
\newcommand{\etal}{et al.}

\newcommand{\mpr}{m_{\rm p}}
\newcommand{\me}{m_{\rm e}}
\newcommand{\Ye}{Y_{\rm e}}
\newcommand{\sigth}{\sigma_{\rm T}}
\newcommand{\alphaf}{\alpha_{\rm F}}
\newcommand{\vecE}{{\bmath E}}
\newcommand{\vecB}{{\bmath B}}
\newcommand{\vechatB}{\hat{\bmath B}}
\newcommand{\veck}{{\bmath k}}
\newcommand{\vechatk}{\hat{\bmath k}}
\newcommand{\vechaty}{\hat{\bmath y}}
\newcommand{\vechatz}{\hat{\bmath z}}
\newcommand{\vechatzp}{\hat{\bmath z}'}
\newcommand{\vece}{{\bmath e}}
\newcommand{\vecr}{{\bmath r}}
\newcommand{\kabs}{\kappa^{\rm abs}}
\newcommand{\ksc}{\kappa^{\rm sc}}

\newcommand{\kffj}{\kappa^{\rm ff,e}_j}
\newcommand{\kiffj}{\kappa^{\rm ff,i}_j}
\newcommand{\kabsj}{\kappa^{\rm abs}_j}
\newcommand{\kscj}{\kappa^{\rm sc}_j}
\newcommand{\kesj}{\kappa^{\rm es}_j}
\newcommand{\kesji}{\kappa^{\rm es}_{ji}}
\newcommand{\kisji}{\kappa^{\rm is}_{ji}}
\newcommand{\kisj}{\kappa^{\rm is}_j}
\newcommand{\keso}{\kappa^{\rm es}_0}

\newcommand{\gff}{\bar{g}^{\rm ff}}
\newcommand{\nel}{n_{\rm e}}
\newcommand{\nion}{n_{\rm i}}
\newcommand{\Teff}{T_{\rm eff}}
\newcommand{\Ebe}{E_{Be}}
\newcommand{\omegab}{\omega_{Be}}
\newcommand{\Ebi}{E_{Bi}}
\newcommand{\omegabi}{\omega_{Bi}}
\newcommand{\Epe}{E_{{\rm p}e}}
\newcommand{\omegap}{\omega_{{\rm p}e}}

\newcommand{\Bq}{B_{\rm Q}}
\newcommand{\thetab}{\theta_B}
\newcommand{\Thetab}{\Theta_B}
\newcommand{\polarb}{\beta}
\newcommand{\uel}{u_{\rm e}}
\newcommand{\uion}{u_{\rm i}}
\newcommand{\vel}{v_{\rm e}}
\newcommand{\vion}{v_{\rm i}}
\newcommand{\nudamp}{\nu_{\rm e}}
\newcommand{\nudampi}{\nu_{\rm i}}
\newcommand{\nudampre}{\nu_{\rm r,e}}
\newcommand{\nudampri}{\nu_{\rm r,i}}
\newcommand{\nudampce}{\nu_{\rm c,e}}
\newcommand{\nudampci}{\nu_{\rm c,i}}
\newcommand{\gamdampi}{\gamma_{\rm i}}
\newcommand{\gamdampre}{\gamma_{\rm r}}
\newcommand{\gamdampce}{\gamma_{\alpha}}

\newcommand{\opbrace}{\xi}
\newcommand{\lambdad}{\lambda_{s}}
\newcommand{\omegabo}{\omega_{Bs}}
\newcommand{\omegapo}{\omega_{{\rm p}s}}

\newcommand{\vecD}{{\bmath D}}

\newcommand{\vecone}{{\bf I}}
\newcommand{\veceps}{{\bepsilon}}
\newcommand{\vecepspl}{\bepsilon^{\rm (p)}}
\newcommand{\vecepsvp}{\Delta\bepsilon^{\rm (v)}}

\newcommand{\epsvp}{\Delta\epsilon^{\rm (v)}}
\newcommand{\vpa}{a}
\newcommand{\vpatwo}{\hat{a}}
\newcommand{\vpq}{q}
\newcommand{\vpm}{m}
\newcommand{\vpr}{r}
\newcommand{\vprtwo}{\hat{r}}
\newcommand{\vecBw}{{\bmath B}_{\rm w}}
\newcommand{\vecHw}{{\bmath H}_{\rm w}}

\newcommand{\deltavp}{\delta_{\rm V}}
\newcommand{\Evp}{E_{\rm V}}
\newcommand{\Evpo}{E_{\rm V}^{(0)}}
\newcommand{\densvp}{\rho_{\rm V}}
\newcommand{\polarbvp}{\beta_{\rm V}}
\newcommand{\holai}{Paper I}
\newcommand{\laiho}{Paper II}
\newcommand{\Gammavp}{\Gamma_{\rm V}}
\newcommand{\taud}{\tau_{\rm d}}
\newcommand{\rhod}{\rho_{\rm d}}
\newcommand{\tauo}{\tau_L}
\newcommand{\kaph}{\kappa_{\rm H}}
\newcommand{\kapl}{\kappa_{\rm L}}
\newcommand{\kappaE}{\kappa_{\rm E}}

\newcommand{\polarbmcp}{\tilde{\beta}}
\newcommand{\gamdampe}{\gamma_{\rm e}}
\newcommand{\Emcp}{E_{\rm Far}}
\newcommand{\thetamcp}{\theta_{\rm coll}}

\title[Atmospheres and Spectra of Strongly Magnetized Neutron Stars II:
Vacuum Polarization]{Atmospheres and Spectra of Strongly
Magnetized Neutron Stars II: Effect of Vacuum Polarization}
\author[W.C.G. Ho and D. Lai]{Wynn C. G. Ho and Dong Lai \\
Center for Radiophysics and Space Research, 
Department of Astronomy, Cornell University
Ithaca, NY 14853, USA \\
{\rm E-mail: wynnho@astro.cornell.edu, dong@astro.cornell.edu}}
\date{Accepted 2002 xxx,
      Received 2002 xxx;
      in original form 2002 xxx}

\pagerange{\pageref{firstpage}--\pageref{lastpage}}

\begin{document}

\maketitle

\label{firstpage}

\begin{abstract}
We study the effect of vacuum polarization on the atmosphere
structure and radiation spectra of neutron stars with surface
magnetic fields $B=10^{14}-10^{15}$~G, as appropriate for
magnetars.
Vacuum polarization modifies the dielectric property of the
medium and gives rise to a resonance feature in the opacity;
this feature is narrow and occurs at a photon energy that
depends on the plasma density.  Vacuum polarization can also
induce resonant conversion of photon modes via a mechanism
analogous to the MSW mechanism for neutrino oscillation.
We construct atmosphere models in radiative equilibrium with
an effective temperature of a few $\times 10^6$~K by solving
the full radiative transfer equations for both polarization
modes in a fully ionized hydrogen plasma.
We discuss the subtleties in treating the vacuum
polarization effects in the atmosphere models and present
approximate solutions to the radiative transfer problem which
bracket the true answer.
We show from both analytic considerations and numerical
calculations that vacuum polarization produces a broad
depression in the X-ray flux at high energies
($\mbox{a few keV}\la E\la\mbox{a few tens of keV}$)
as compared to models without vacuum polarization; this arises
from the density dependence of the vacuum resonance feature and
the large density gradient present in the atmosphere.
Thus the vacuum polarization effect softens the high energy tail
of the thermal spectrum, although the atmospheric emission is
still harder than the blackbody spectrum because of the non-grey
opacities.
We also show that the depression of continuum flux
strongly suppresses the equivalent width of
the ion cyclotron line and therefore makes the line more difficult
to observe.
\end{abstract}


\setcounter{equation}{0}
\section{Introduction} \label{sec:intro}

Thermal radiation from the surface of isolated neutron stars (NSs)
can provide invaluable information on the physical properties and
evolution of NSs.  In the last few years, such radiation has been
detected in four types of isolated NSs: (1) radio pulsars (see Becker~2000),
(2) old, radio-quiet NSs (not associated with supernova
remnants), some of which may be accreting from the interstellar
medium (see Treves \etal~2000),
(3) soft gamma-ray repeaters (SGRs) and anomalous X-ray pulsars (AXPs)
(see Hurley~2000; Israel, Mereghetti, \& Stella~2001; Mereghetti~2001;
Perna \etal~2001), which form a potentially new class of NSs
(``magnetars'') endowed with superstrong ($B\ga 10^{14}$~G) magnetic
fields (see Thompson \& Duncan~1996; Thompson~2001),
and (4) young, radio-quiet NSs in supernova remnants, which may
include AXPs or SGRs (see Pavlov \etal~2001).
The NS surface emission is
mediated by the thin atmospheric layer (with scale height
$\sim 0.1-10$~cm and density $\sim 0.1-100$~g/cm$^3$) that
covers the stellar surface.  Therefore, to properly interpret
the observations of NS surface emission and to provide
accurate constraints on the physical properties of NSs, it is
important to understand in detail the radiative properties
of NS atmospheres in the presence of strong magnetic fields
(see Pavlov \etal~1995 and Ho \& Lai~2001 for more detailed
references on observations and on previous works of NS
atmosphere modeling).

This paper is the second in a series where we systematically
investigate the atmosphere and spectra of strongly magnetized NSs.
In Ho \& Lai~(2001, hereafter \holai), we constructed
self-consistent NS atmosphere models in radiative equilibrium
with magnetic field $B\sim 10^{12}-10^{15}$~G and $\Teff\sim 10^6-10^7$~K
and assuming the atmosphere is composed of fully ionized hydrogen
or helium.  The radiative opacities include
free-free absorption and scattering by both electrons and ions
computed for the two photon polarization modes in the magnetized
electron-ion plasma.  It was found that, in general, the emergent
thermal radiation exhibits significant deviation from blackbody,
with harder spectra at high energies; the spectra can also show a
broad feature ($\Delta E/\Ebi\sim 1$) around the ion cyclotron
resonance $\Ebi=0.63\,(Z/A)(B/10^{14}\mbox{ G})$~keV, where $Z$ and $A$
are the atomic charge and atomic mass of the ion, respectively
(see also Zane \etal~2001), and this feature is particularly
pronounced when $\Ebi\ga 3k\Teff$
(however, as we show in this paper, vacuum polarization can significantly
diminish the equivalent width of the ion cyclotron feature;
see Sections~\ref{sec:densd} and \ref{sec:results}).

In this paper, we study the effect of vacuum polarization on
the atmosphere structure and radiation spectra of strongly
magnetized NSs.  It is well-known that polarization
of the vacuum due to virtual $e^+e^-$ pairs becomes significant
when $B\ga\Bq$, where $\Bq=\me^2c^3/e\hbar=4.414\times 10^{13}$~G
is the magnetic field at which the electron cyclotron energy
$\hbar\omegab=\hbar eB/\me c$ equals $\me c^2$.  Vacuum
polarization modifies the dielectric property of the medium
and the polarization of photon modes
(e.g., Adler~1971; Tsai \& Erber~1975; Gnedin, Pavlov, \& Shibanov~1978;
Heyl \& Hernquist~1997b), thereby altering the
radiative scattering and absorption opacities 
(e.g., M\'{e}sz\'{a}ros \& Ventura~1979; Pavlov \& Gnedin~1984;
see M\'{e}sz\'{a}ros~1992 for review).
Of particular interest is the ``vacuum resonance'' phenomenon, which
occurs when the effects of the vacuum and plasma on the linear
polarization of the modes cancel each other, giving rise to
``resonant'' features in the radiative opacities
(e.g., Pavlov \& Shibanov~1979; Ventura, Nagel, \& M\'{e}sz\'{a}ros~1979;
Bulik \& Miller~1997).
At a given density $\rho$, the vacuum-induced
resonance is located at the photon energy
\be
\Evp \approx 1.02\lp\frac{\Ye\rho}{\mbox{1 g cm$^{-3}$}}\rp^{1/2}
 \lp\frac{B}{\mbox{10$^{14}$ G}}\rp^{-1}f(B)\mbox{ keV},
\ee
where $\Ye=Z/A$ is the electron fraction and $f(B)$ is a
slowly-varying function of $B$ [$f(B)=1$ for $B\ll\Bq$ and is of
order a few for $B\sim 10^{14}-10^{16}$~G; see Section~\ref{sec:polar}].
It has been suggested that
this vacuum resonance may manifest itself as absorption-like
features in the spectra of X-ray pulsars and magnetars
(e.g., Ventura, Nagel, \& M\'{e}sz\'{a}ros~1979; Bulik \& Miller~1997).
However, as we show both analytically and numerically in this paper
(see also Bezchastnov \etal~1996),
since the atmosphere spans a wide range of densities and $\Evp$
depends on $\rho$, the spectral feature associated
with $\Evp$ is spread out significantly and manifests as a broad
``depression'' in the radiation spectrum.

Furthermore, in a recent paper (Lai \& Ho~2002a, hereafter \laiho),
we study the effect of the mode conversion that is associated
with the vacuum resonance (see also Gnedin \etal~1978; Pavlov \&
Gnedin~1984): a photon propagating outward in the NS atmosphere can convert
from one polarization mode into another as it traverses the resonant
density $\densvp$, which is given by
\be
\densvp \approx 0.96\,\Ye^{-1}\lp\frac{E}{\mbox{1 keV}}\rp^2
 \lp\frac{B}{\mbox{10$^{14}$ G}}\rp^2f(B)^{-2}\mbox{ g cm$^{-3}$}.
\ee
(Across the resonance, the orientation of the polarization ellipse
rotates by $90^\circ$, although the helicity does not change.)
This resonant mode conversion is analogous to the
Mikheyev-Smirnov-Wolfenstein (MSW) effect for neutrino oscillations
(e.g., Bahcall~1989; Haxton~1995) and is effective
for $E\ga$~a~few~keV (for which the propagation is adiabatic) and
ineffective for $E\la$~a~few~keV (the adiabatic condition breaks down).
Because the two photon modes have vastly different opacities,
this vacuum-induced mode conversion can significantly affect
radiative transfer in magnetar atmospheres (see Section~\ref{sec:modeconv}).
To properly account for this mode conversion effect in the atmosphere
models, one must go beyond the modal description of the radiation
field by formulating and solving the transfer equation in terms of the
photon density matrix and including the effect of the refractive
index.  This is beyond the scope of this paper.  Instead, we consider
two limiting cases in this paper: no mode conversion and complete
mode conversion.  These two limits correspond to different
ways of labeling the photon modes, and the resulting model
atmosphere spectra should approximately bracket the true spectra.

There have been few previous works studying the effect of
vacuum polarization on the radiation spectra of strongly magnetized NSs.
Bezchastnov \etal~(1996) reported NS atmosphere models
with $B\sim 10^{14}-10^{16}$~G and $\Teff\approx 10^8$~K.
This temperature is not appropriate for SGRs and AXPs, which have
$\Teff\sim\mbox{ a few}\times 10^6$~K (see, e.g., Mereghetti~2001;
Perna \etal~2001).
Although the vacuum-induced mode conversion was not discussed,
the labeling of the photon modes in
Bezchastnov \etal~(1996) amounts to assuming complete mode conversion
(see Section~\ref{sec:modeconv}).
Bulik \& Miller~(1997) considered how blackbody radiation with
$T\approx 10^8$~K is modified as it passes through a tenuous
plasma where the opacity is affected by vacuum polarization,
but no self-consistent atmosphere modeling, which is needed to
determine the density and temperature profiles, was attempted,
and mode conversion was neglected.
\"{O}zel~(2001) studied magnetar atmosphere models
(with $B\ga 10^{14}$~G and $\Teff\sim\mbox{ a few}\times 10^6$~K)
which include the effect of vacuum polarization on the opacities
(see also Shibanov \etal~1992 for $B\sim 10^{12}$~G
atmosphere models which include a similar vacuum polarization effect
on the opacities).
Besides neglecting resonant mode conversion, \"{O}zel's work
also neglects the ion effect and adopts a ``saturation'' scheme
to smooth out the vacuum resonance feature in the opacity.
Our analytical and numerical
consideration of the vacuum resonance feature shows that such
a ``saturation'' scheme cannot be justified
(see Sections~\ref{sec:opacity} and \ref{sec:saturation}).
Indeed, as we discuss in this paper, even when neglecting resonant
mode conversion, great care must be taken to properly handle
the narrow and density-dependent vacuum-induced opacity feature
in the radiative transfer (see Section~\ref{sec:numcomp}).

In Section~\ref{sec:polar}, we study the effect of vacuum polarization
on the photon propagation modes in a magnetized electron-ion plasma
and clarify the nature of the vacuum resonance and the associated mode
conversion phenomenon.
In Section~\ref{sec:opacity}, we discuss the change in the radiative
opacities as a result of vacuum polarization.
Section~\ref{sec:densd} shows the qualitative effect of the vacuum
resonance feature on the NS surface emission.
The difficulties encountered in solving the radiative
transfer equation caused by vacuum polarization is discussed
in Section~\ref{sec:numcomp}.
We present atmosphere models and spectra for different magnetic
field strengths in Section~\ref{sec:results}.
Section~\ref{sec:discussion} summarizes and discusses the
implications of our results.
Several ``toy'' atmosphere models that include opacities which mimic
the vacuum polarization effects are discussed in Appendix~\ref{sec:toy}.

\setcounter{equation}{0}
\section{Photon Polarization Modes in a Magnetized Electron-Ion Plasma
Including Vacuum Polarization}
\label{sec:polar}

In this section, we outline the derivation of the photon polarization
modes in a magnetized electron-ion plasma including the effect of
vacuum polarization.
Following the standard convention, the two normal modes are termed
the extraordinary mode (X-mode, $j = 1$), which is mostly polarized
perpendicular to the
$\veck$-$\vecB$ plane, and the ordinary mode (O-mode, $j = 2$), which
is mostly polarized parallel to the $\veck$-$\vecB$ plane,
where $\veck$ is the wave vector along the wave propagation
direction and $\vecB$ is the external magnetic field
(e.g., M\'{e}sz\'{a}ros 1992).
We shall see that this classification of the modes becomes ambiguous
near the vacuum resonance, where mode conversion can occur
(Sections~\ref{sec:vpres} and \ref{sec:modeconv}).

\subsection{Dielectric and Permeability Tensors}
\label{sec:dielectric}

For a cold electron-ion plasma, the plasma contribution to the
dielectric tensor $\vecepspl$ can be written, in the coordinate
system $x'y'z'$ with $\vecB$ along $z'$, as (e.g., Shafranov~1967)
\be
\lb\vecepspl\rb_{\vechatzp=\vechatB}= \lb \begin{array}{ccc}
\varepsilon & ig & 0 \\
-ig & \varepsilon & 0 \\
0 & 0 & \eta
\end{array} \rb, \label{eq:epsij0}
\ee
where
\ba
\varepsilon & = & 1-\sum_s\frac{\lambdad v_s}{\lambdad^2 - u_s}
 \approx 1-\vel\frac{1-M\uion}{(1-\uion)(1-\uel)} \label{eq:epsii0} \\
\eta & = & 1-\sum_s\frac{v_s}{\lambdad} \approx 1 - \vel \label{eq:epszz0} \\
g & = & -\sum_s\frac{\mbox{sign}(q_s)u_s^{1/2}v_s}{\lambdad^2-u_s}
 \approx \frac{\vel\uel^{1/2}}{(1-\uion)(1-\uel)}.  \label{eq:epsxy0}
\ea
In equations~(\ref{eq:epsii0})-(\ref{eq:epsxy0}), the sums run
over each charged particle species $s$ (electron and ion) in the plasma,
and $u_s = \omegabo^2/\omega^2$ and $v_s = \omegapo^2/\omega^2$, where
$\omegabo=|q_s|B/(m_s c)$ is the cyclotron frequency and
$\omegapo=(4\pi n_sq_s^2/m_s)^{1/2}$ is the plasma frequency
of charged particle $s$.
Damping of the particle motion is accounted for by
$\lambdad = 1+i\nu_s/\omega$, where $\nu_s$ is the damping rate.
In the second equalities of equations~(\ref{eq:epsii0})-(\ref{eq:epsxy0}),
we have taken the $M\equiv A\mpr/(Z\me)=\vel/\vion=(\uel/\uion)^{1/2}\gg 1$
limit and assumed small damping ($\nu_s\ll\omega$ or
$\lambdad\rightarrow 1$).  The relevant dimensionless parameters are
\be
\uel = \frac{\omegab^2}{\omega^2}, \qquad
\uion = \frac{\omegabi^2}{\omega^2}, \qquad \vel = \frac{\omegap^2}{\omega^2},
\ee
where the electron cyclotron frequency
$\omegab$, the ion cyclotron frequency $\omegabi$, and the electron
plasma frequency $\omegap$ are given by
\be
\Ebe = \hbar\omegab = \hbar\frac{eB}{\me c}
 = 1.158\,B_{14}\mbox{ MeV}
\ee
\be
\Ebi = \hbar\omegabi = \hbar\frac{ZeB}{A\mpr c}
 = 0.6305\,B_{14}\lp\frac{Z}{A}\rp\mbox{ keV} \label{eq:ioncycen}
\ee
\be
\Epe = \hbar\omegap = \hbar\lp\frac{4\pi e^2\nel}{\me}\rp^{1/2}
 \!\!\!\!\!\!  = 28.71\,\lp\frac{Z}{A}\rp^{1/2}\!\!\!\rho_1^{1/2}\mbox{ eV},
\ee
respectively, $B_{14}=B/$(10$^{14}$~G), and $\rho_1=\rho/$(1~g~cm$^{-3}$).

Vacuum polarization contributes a correction to the dielectric tensor:
\be
\vecepsvp = \lp\vpa - 1\rp\vecone + \vpq\vechatB\vechatB,
\ee
where $\vecone$ is the unit tensor and
$\vechatB=\vecB/B$ is the unit vector along $\vecB$.
The magnetic permeability tensor $\bmu$ also deviates from unity
because of vacuum polarization; the magnetic field $\vecHw$
of an electromagnetic wave is related to its magnetic induction
$\vecBw$ by
\be
\vecHw=\bmu^{-1}\cdot\vecBw=\lp\vpa\vecone+\vpm\vechatB\vechatB\rp\cdot\vecBw.
 \label{eq:permeab}
\ee
For $\hbar\omega\ll\me c^2$, general expressions for the vacuum
polarization coefficients $\vpa$, $\vpq$, and $\vpm$ are given
in Heyl \& Hernquist~(1997a).
In the weak-field limit, $B\ll\Bq=4.414\times 10^{13}$~G,
they are given by
\ba
\vpa & = & 1-2\deltavp \label{eq:vpalo} \\
\vpq & = & 7\deltavp \label{eq:vpqlo} \\
\vpm & = & -4\deltavp, \label{eq:vpmlo}
\ea
where
\be
\deltavp = \frac{\alphaf}{45\pi}b^2, \label{eq:deltavp}
\ee
$\alphaf=e^2/\hbar c=1/137$ is the fine structure constant and
$b = B/\Bq$ (Adler~1971).
For $B\ga\Bq$, we use the expansions given in Heyl \& Hernquist~(1997a,b;
see also Tsai \& Erber~1975) to find
\ba
\vpa & \approx & 1+\frac{\alphaf}{2\pi}\lb 1.195-\frac{2}{3}\ln b
 -\frac{1}{b}\lp 0.8553+\ln b\rp \right. \nonumber \\
 && \qquad \qquad \left. -\frac{1}{2b^2}\rb \label{eq:vpahi} \\
\vpq & \approx & -\frac{\alphaf}{2\pi} \lb
 -\frac{2}{3}b + 1.272 - \frac{1}{b}\lp 0.3070+\ln b\rp \right. \nonumber \\
 && \qquad \qquad \left. -0.7003\frac{1}{b^2}\rb \label{eq:vpqhi} \\
\vpm & \approx & -\frac{\alphaf}{2\pi} \lb\frac{2}{3}+\frac{1}{b}\lp 0.1447-\ln b\rp
 -\frac{1}{b^2}\rb.  \label{eq:vpmhi}
\ea

When $|\epsvp_{ij}|\ll 1$ or $b\ll 3\pi/\alphaf$
($B\ll 5\times10^{16}$~G), the plasma and vacuum contributions to
the dielectric tensor can be added linearly, i.e.,
$\veceps=\vecepspl+\vecepsvp$.
In the frame with $\vechatB$ along $\vechatzp$,
\ba
\lb\veceps\rb_{\vechatzp=\vechatB} & = & \lb \begin{array}{ccc}
\varepsilon & ig & 0 \\
-ig & \varepsilon & 0 \\
0 & 0 & \eta
\end{array} \rb
+ \lb \begin{array}{ccc}
\vpatwo & 0 & 0 \\
0 & \vpatwo & 0 \\
0 & 0 & \vpatwo+\vpq
\end{array} \rb \nonumber \\
 & = & \lb \begin{array}{ccc}
\varepsilon' & ig & 0 \\
-ig & \varepsilon' & 0 \\
0 & 0 & \eta'
\end{array} \rb, \label{eq:epsij}
\ea
where $\vpatwo=\vpa-1$, $\varepsilon'=\varepsilon+\vpatwo$, and
$\eta'=\eta+\vpatwo+\vpq$.  The total magnetic permeability
is given by equation~(\ref{eq:permeab}).

\subsection{Equations for the Polarization Modes}
\label{sec:modes}

Using the electric displacement $\vecD=\veceps\cdot\vecE$
and equation~(\ref{eq:permeab})
in the Maxwell equations, we obtain the equation for
plane waves with $\vecE\propto e^{i(\veck\cdot\vecr-\omega t)}$:
\be
\left\{ \frac{1}{\vpa}\epsilon_{ij}+n^2\lb\hat{k}_i\hat{k}_j-\delta_{ij}
 - \frac{\vpm}{\vpa}(\hat{k}\times\hat{B})_i(\hat{k}\times\hat{B})_j
 \rb \right\} E_j=0, \label{eq:nrefract}
\ee
where $n=ck/\omega$ is the refractive index and $\vechatk=\veck/k$.
In the coordinate system $xyz$ with $\veck$ along the $z$-axis and
$\vecB$ in the $x$-$z$ plane, such that
$\vechatk\times\vechatB=-\sin\thetab\vechaty$ ($\thetab$ is the
angle between $\vechatk$ and $\vechatB$), the dielectric tensor
[eq.~(\ref{eq:epsij})] is given by
\ba
\lb\veceps\rb_{\vechatz=\vechatk} & = & \lb \begin{array}{cc}
\varepsilon'\cos^2\thetab+\eta'\sin^2\thetab & ig\cos\thetab \\
-ig\cos\thetab & \varepsilon' \\
\lp\varepsilon'-\eta'\rp\sin\thetab\cos\thetab & ig\sin\thetab \\
\end{array} \right. \nonumber \\
 && \qquad \qquad \left. \begin{array}{c}
 \lp\varepsilon'-\eta'\rp\sin\thetab\cos\thetab \\
 -ig\sin\thetab \\
 \varepsilon'\sin^2\thetab+\eta'\cos^2\thetab
\end{array}\rb. \label{eq:epsijx}
\ea
The $z$-component of equation~({\ref{eq:nrefract}) gives
\be
E_z = -\epsilon_{zz}^{-1}\lp\epsilon_{zx}E_x+\epsilon_{zy}E_y\rp.
\label{eq:EzExEy}
\ee
Reinserting this back into equation~(\ref{eq:nrefract}) yields
\ba
\lp \begin{array}{cc}
\eta_{xx} - n^2 & \eta_{xy} \\
\eta_{yx} & \eta_{yy} - \vpr n^2
\end{array} \rp
\lp \begin{array}{c} E_x \\ E_y \end{array} \rp = 0, \label{eq:nrefractmatrix}
\ea
where $\vpr = 1+\vprtwo \equiv 1 + (\vpm/\vpa)\sin^2\thetab$ and
\ba
\eta_{xx} & = & \frac{1}{\vpa\epsilon_{zz}}\lp\epsilon_{zz}\epsilon_{xx}
 -\epsilon_{xz}\epsilon_{zx}\rp = \frac{1}{\vpa\epsilon_{zz}}
 \varepsilon'\eta' \\
\eta_{yy} & = & \frac{1}{\vpa\epsilon_{zz}}\lp\epsilon_{zz}\epsilon_{yy}
 -\epsilon_{yz}\epsilon_{zy}\rp \nonumber \\
 & = & \frac{1}{\vpa\epsilon_{zz}} \lb\lp
 \varepsilon'^2-g^2-\varepsilon'\eta'\rp\sin^2\thetab+\varepsilon'\eta'\rb \\
\eta_{yx} & = & -\eta_{xy} = \frac{1}{\vpa\epsilon_{zz}}
 \lp\epsilon_{zz}\epsilon_{yx}-\epsilon_{yz}\epsilon_{zx}\rp \nonumber \\
 & = & \frac{1}{\vpa\epsilon_{zz}}\lp -ig\eta'\cos\thetab\rp.
\ea
We write the unit polarization vector as
\be
\vecE=\vece^j=\frac{1}{(1+K_j^2+K_{z,j}^2)^{1/2}}(iK_j,1,iK_{z,j}),
\ee
where $iK_j=E_x/E_y$, $iK_{z,j}=E_z/E_y$, and $j$ is the mode index
($j=1$ for the X-mode, and $j=2$ for the O-mode).
Eliminating $n^2$ from equation~(\ref{eq:nrefractmatrix}), we obtain
\be
K_j = \polarb\lb 1+(-1)^j\lp 1+\frac{\vpr}{\polarb^2}\rp^{1/2}\rb.
 \label{eq:polark}
\ee
The polarization parameter $\polarb$ is given by
\ba
\polarb & = & -i\frac{\vpr\eta_{xx}-\eta_{yy}}{2\eta_{yx}} \nonumber \\
 & = & -\frac{\lp\varepsilon'^2-g^2-\varepsilon'\eta'\rp\sin^2\thetab
 +\varepsilon'\eta'(1-\vpr)}{2g\eta'\cos\thetab}. \label{eq:polarb1}
\ea
Using $\varepsilon'=\varepsilon+\vpatwo$ and $\eta'=\eta+\vpatwo+\vpq$
[see eq.~(\ref{eq:epsij})] and $1-\vpr=-(\vpm/\vpa)\sin^2\thetab$,
we write
\be
\polarb = \polarb_0\polarbvp, \label{eq:polarb}
\ee
where $\polarb_0$ is the polarization parameter in the absence of
vacuum polarization
\ba
\polarb_0 & = & -\frac{\lp\varepsilon^2-g^2
 -\varepsilon\eta\rp\sin^2\thetab}{2g\eta\cos\thetab} \nonumber \\
 & = & \frac{\uel^{1/2}}{2(1-\vel)}\frac{\sin^2\!\thetab}{\cos\thetab}
  \lp 1-\uion-\frac{ 1+\vel}{M}\rp \label{eq:polarb0}
\ea
and $\polarbvp$ is the correction factor due to vacuum polarization
\ba
\polarbvp & = & \lp 1+\frac{\vpatwo+\vpq}{\eta}\rp^{-1}
\lb 1+\frac{\varepsilon(\vpatwo-\vpq)-\eta\vpatwo-\varepsilon\eta\vpm
}{\varepsilon^2-g^2-\varepsilon\eta}\rb \nonumber \\
 & = & \lp 1+\frac{\vpatwo+\vpq}{1-\vel}\rp^{-1}
\left\{ 1+ \frac{\lp\vpq+\vpm\rp\lp 1-\uel\rp}{\uel\vel} \right. \nonumber
\ea
\be
 \times \left.
\lb\frac{\lp 1-\uion\rp\lp 1-\frac{\vpatwo+\vpm}{\vpq+\vpm}\vel\rp
 - \frac{\vel\lp 1-M\uion\rp}{1-\uel}
 \frac{-\vpatwo+\vpq+\vpm(1-\vel)}{\vpq+\vpm}}{1-\uion-M^{-1}\lp 1+\vel\rp}\rb
\right\}. \label{eq:polarbvp}
\ee
For $\vel\ll 1$, equation~(\ref{eq:polarbvp}) simplifies to
\be
\polarbvp \approx 1+\frac{\lp\vpq+\vpm\rp\lp 1-\uel\rp}{\uel\vel}.
 \label{eq:polarbvp0}
\ee
In addition, from equation~(\ref{eq:EzExEy}), we find
\ba
K_{z,j} & = & -\frac{\lp\varepsilon-\eta-\vpq\rp\sin\thetab\cos\thetab K_j
 +g\sin\thetab}{\varepsilon\sin^2\thetab+\lp\eta+\vpq\rp\cos^2\thetab+\vpatwo}
 \nonumber \\
 & = & \lb\uel\vel\lp 1-\uion-\frac{1}{M}\rp \sin\thetab\cos\thetab K_j
 \right. \nonumber
\ea
\begin{displaymath}
 \qquad \times \left. \lp 1+q\frac{1-\uel}{\uel\vel}\frac{1-\uion}{1-\uion-M^{-1}}
 \rp-\uel^{1/2}\vel\sin\thetab \rb
\end{displaymath}
\begin{displaymath}
 \qquad \times \left\{ \lp 1-\uel\rp \lp 1-\uion\rp
 \lp 1+\vpatwo+\vpq\cos^2\thetab\rp \right.
\end{displaymath}
\be
 \qquad \left. -\vel\lb\lp 1-\uion\rp\lp 1-\uel\cos^2\thetab\rp
 - M\uion\sin^2\thetab\rb \right\}^{-1} .  \label{eq:polarkz}
\ee
It is evident from equation~(\ref{eq:polarkz}) that
the component of $\vece^j$ along $\veck$ is of order
$\vel\propto\rho/\omega^2$, and thus, at sufficiently low
densities, $K_{z,j}$ can be neglected so that the modes
are transverse.
 
Finally, the refractive index $n_j$ of the mode can be obtained from
equation~(\ref{eq:nrefractmatrix}), which gives
\be
n_j^2 = \frac{g\eta'}{\vpa\epsilon_{zz}}\lp\frac{\varepsilon'}{g}
 + \frac{1}{K_j}\cos\thetab\rp, \label{eq:nrefract2}
\ee
where $\epsilon_{zz}=\varepsilon'\sin^2\thetab+\eta'\cos^2\thetab$
[see eq.~(\ref{eq:epsijx})] and $\varepsilon'$, $\eta'$, $g$, and $\vpa$
are as given in equation~(\ref{eq:epsij}).

\subsection{Vacuum Resonance}
\label{sec:vpres}

The polarization parameter $\polarb$
[eqs.~(\ref{eq:polarb1})-(\ref{eq:polarbvp0})] directly determines the
characteristics of photon normal modes in the medium.
For most energies
(away from the resonance points, where $\polarb=0$; see
below), $|\polarb|\propto\uel^{1/2}\gg 1$, and we have
\be
K_1 \approx -\frac{\vpr}{2\polarb} \approx -\frac{1}{2\polarb},
 \qquad K_2 \approx 2\polarb \qquad (|\polarb|\gg 1). \label{eq:Kjhigh}
\ee
Thus the extraordinary mode ($j=1$) is almost linearly polarized
with its electric field perpendicular to the $\veck$-$\vecB$ plane,
while the electric field of the ordinary mode ($j=2$) is in the
$\veck$-$\vecB$ plane.

The condition $\polarb=0$ specifies the resonance points.
For $|\polarb|\ll 1$, equation~(\ref{eq:polark}) yields
\be
K_j \approx \mbox{sign}(\polarb)\lb\lp -1\rp^j\vpr^{1/2}+|\polarb|\rb
 \qquad (|\polarb|\ll 1). \label{eq:polarklo}
\ee
Thus $|K_j|\approx 1$ at $\polarb=0$, and both modes are circularly
polarized.
In general, at a given density $\rho$, there are three critical photon
energies at which $\polarb=0$.  The first is located at
$\omega/\omegap\approx 1-(\omegap/\omegab)^2/[2(\vpq+\vpm)]$,
i.e., close to $\omega=\omegab$.  We will ignore this critical energy
since we are considering only the $\omega\ll\omegab$ regime in this
paper.
The second critical energy is at $\omega=\omegabi$; this
is simply the ion cyclotron resonance and is unrelated to vacuum
polarization.  The third critical point (``vacuum resonance'')
is located at [see eqs.~(\ref{eq:polarbvp}) and (\ref{eq:polarbvp0})]
\be
\vel \approx \lp\vpq+\vpm\rp\lp 1-\frac{1}{\uel}\rp\lp 1-\vpatwo-\vpm\rp
 \label{eq:velvp}
\ee
or at energy
\be
\Evp = \frac{\hbar\omegap}{\sqrt{\vpq+\vpm}}\lp 1+\frac{1}{2\uel}
 +\frac{\vpatwo+\vpm}{2}\rp \approx \frac{\hbar\omegap}{\sqrt{\vpq+\vpm}}.
 \label{eq:evp1}
\ee
Note that for $b\ll 3\pi/\alphaf$ ($B\ll 5\times 10^{16}$~G),
$|\vpatwo|$, $|\vpq|$, $|\vpm|\ll 1$, so that equations~(\ref{eq:velvp})
and (\ref{eq:evp1}) are good approximations.
For $B\ll\Bq$, we use equations~(\ref{eq:vpqlo})-(\ref{eq:deltavp})
to obtain
\ba
\Evp(B\ll\Bq) & = & \Evpo = \lp\frac{15\pi}{\alphaf}\rp^{1/2}
 \frac{\omegap}{\omegab}\me c^2 \nonumber \\
 & = & 1.02\,\Ye^{1/2}\rho_1^{1/2} B_{14}^{-1}\mbox{ keV}. \label{eq:evp0}
\ea
For $B\gg\Bq$, we use the leading terms in equations~(\ref{eq:vpqhi}) and
(\ref{eq:vpmhi}) to obtain
\ba
\Evp(B\gg\Bq) & \approx & \lp\frac{3\pi}{\alphaf}\rp^{1/2}\lp\frac{\Bq}{B}\rp^{1/2}\hbar\omegap \nonumber \\
 & = & 0.69\,\Ye^{1/2}\rho_1^{1/2} B_{14}^{-1/2}\mbox{ keV}.  \label{eq:evp}
\ea
For convenience, we define the dimensionless function $f(B)$ via
$\Evp=\Evpo f(B)$, so that
\be
f(B) = \frac{\Evp}{\Evpo} \approx \lp\frac{3\deltavp}{\vpq+\vpm}\rp^{1/2}.
 \label{eq:evplohi}
\ee
Figure~\ref{fig:evp} shows that $f(B)$ is a slowly-varying function of
$b$ [$f(B)=1$ for $b\ll 1$ and
$f(B)\rightarrow (b/5)^{1/2}\approx 0.673\,B_{14}^{1/2}$ for $b\gg 1$;
$f(B)$ varies from $\approx 1$ at $B_{14}=1$ to $6.7$ at $B_{14}=100$].

\begin{figure}
\centering
\includegraphics[height=8cm]{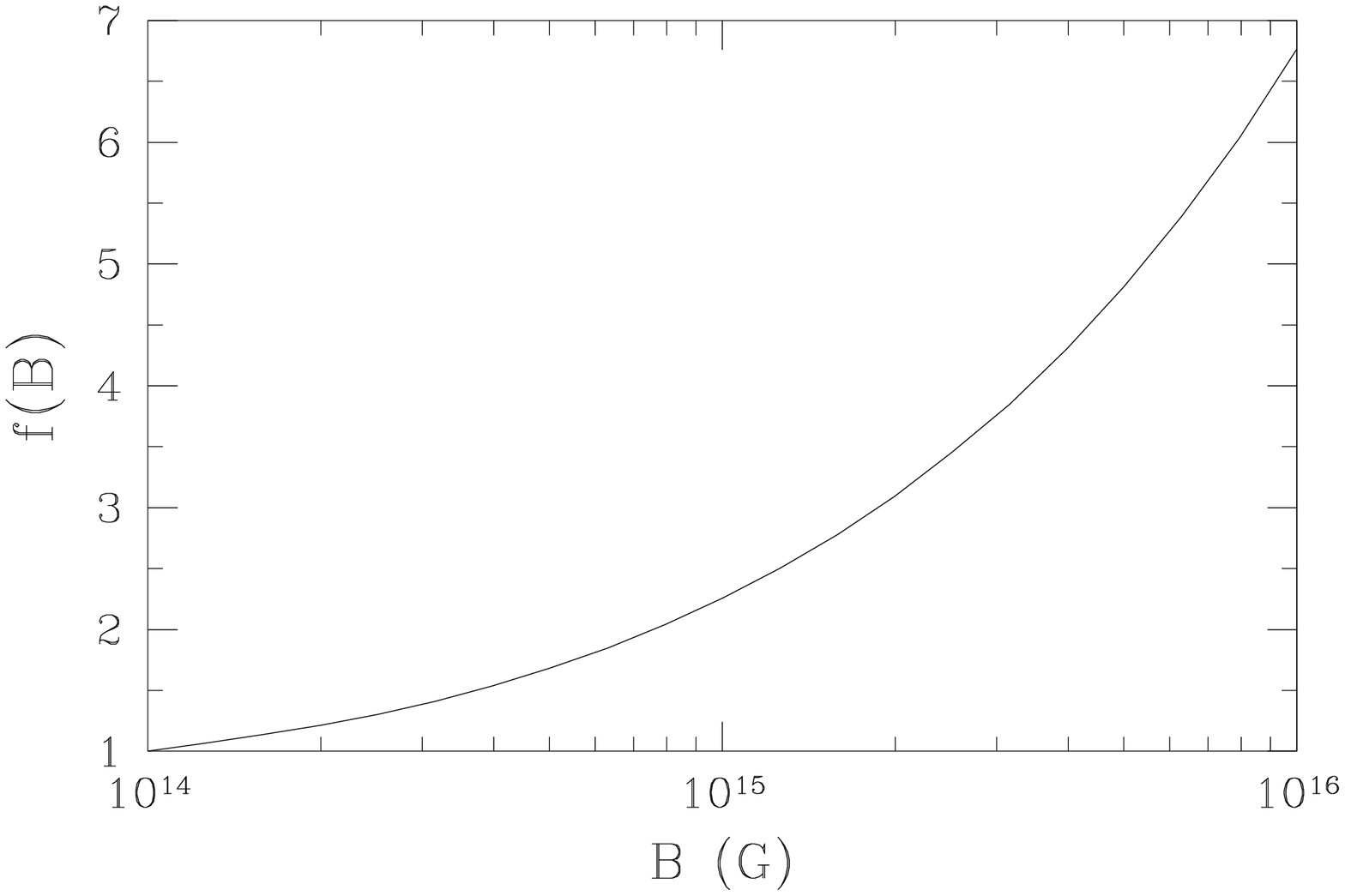}
\caption{
The ratio $f(B)=\Evp(B)/\Evpo(B)$ as a function of $B$, where
$\Evp(B)$ is the vacuum resonance energy [eq.~(\ref{eq:evp1})]
and $\Evpo(B)=\Evp(B\ll\Bq)$ [eq.~(\ref{eq:evp0})].
\label{fig:evp}
}
\end{figure}

Since $\Evp$ depends on density, a photon with a given energy $E$
traveling in an inhomogeneous medium encounters the vacuum
resonance ($\polarb=0$) at the density
\be
\densvp = 0.964\,Y_e^{-1}(B_{14}E_1)^2 f(B)^{-2}\,{\rm g~cm}^{-3},
\label{eq:densvp}
\ee
where $E_1=E/(1~{\rm keV})$.
In Section~\ref{sec:modeconv}, we discuss how the photon polarization
state changes as the photon traverses the resonant density.

\subsection{Resonant Conversion of Photon Modes}
\label{sec:modeconv}

In Section~\ref{sec:modes}, we described the photon modes as the
extraordinary mode (X-mode) and ordinary mode (O-mode).
This standard way of classifying photon modes
(e.g., M\'{e}sz\'{a}ros 1992) is useful when $|\polarb|\gg 1$:
the X-mode has $|K_j|\ll 1$, and its $\vecE$ is
mostly perpendicular to the ${\vechatk}$-$\vechatB$ plane; the
O-mode has $|K_j|\gg 1$ and is polarized along the
$\vechatk$-$\vechatB$ plane.   The advantage of such a classification
is that the X-mode and O-mode interact very differently with matter:
the O-mode opacity is largely unaffected by the magnetic field,
while the X-mode opacity is significantly reduced (by a factor
of order $\omega^2/\omegab^2)$ from the zero-field value
(see Section~\ref{sec:opacity}).
However, the X and O-mode classification becomes ambiguous near
$|\polarb|=0$ ($|K_j|=1$).  Indeed, equation~(\ref{eq:polark}) or
(\ref{eq:polarklo}) shows that $K_j$ is discontinuous (and changes
sign) when crossing through $\polarb=0$ at the resonance density
[eq.~(\ref{eq:densvp})].  This discontinuity in the mode polarization
vector can be avoided by adopting a different mode classification
scheme: instead of equation~(\ref{eq:polark}), we define the
plus-mode and minus-mode according to
\be
K_\pm = \polarb \pm \lp \polarb^2 + \vpr\rp^{1/2}. \label{eq:polarkpm}
\ee
Clearly, $K_\pm$ are continuous functions of $\polarb$; the plus-mode
always has $K_+>0$, and the minus-mode always has $K_-<0$.

\begin{figure}
\centering
\includegraphics[height=8cm]{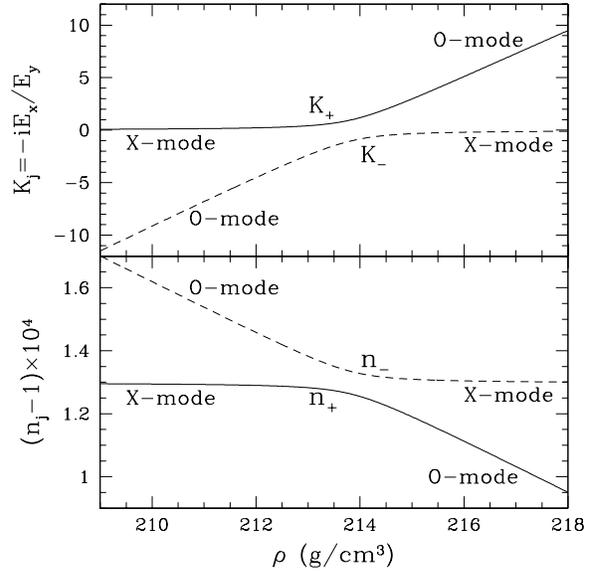}
\caption{
The polarization ellipticities $K_j$ [upper panel; see eq.~(\ref{eq:polarkpm})]
and refractive indices $n_j$ [lower panel; see eq.~(\ref{eq:nrefract2})]
of the photon modes as functions of density near the vacuum resonance for
$B=5\times 10^{14}$~G, $\thetab=45^\circ$, $E=5$~keV, and $\Ye=1$.
\label{fig:polark}
}
\end{figure}

The relationship between the two schemes of mode classification
($j=1,2$ for the X and O-modes, and $j=\pm$ for the plus and minus-modes)
is illustrated by Figure~\ref{fig:polark}, which shows the values
of $K_\pm$ and the refractive indices $n_\pm=ck_\pm/\omega$ for
the plus and minus-modes as a function of density near the vacuum
resonance for $E=5$~keV, $B=5\times 10^{14}$~G, and $\thetab=45^\circ$
(see Fig.~1 in \laiho\ for the $B=10^{14}$~G case).
The plus-mode (minus-mode) manifests as the O-mode
(X-mode) at high densities but becomes the X-mode (O-mode) at low densities.
If the density variation is sufficiently gentle, an O-mode photon
created at high densities will remain on the $K_+$-trajectory
as it travels outward and will adiabatically
convert into the X-mode after traversing the resonance
density.\footnote{In other words, as a photon propagates across the vacuum
resonance under the adiabatic condition, we use the term mode conversion
to imply that the character of the mode, such as the opacity, changes
significantly.  For example, as a high opacity O-mode photon propagates
from high to low density, it becomes a low opacity X-mode photon.
Alternatively, this adiabatic regime could be called mode conservation
since the photon remains on the same plus-mode (or minus-mode) branch
across the resonance.  We use the former terminology.}
This is analogous to the MSW mechanism for neutrino oscillation
(e.g., Bahcall~1989; Haxton~1995).  As shown in \laiho\
(see also Gnedin \etal~1978; Pavlov \& Gnedin~1984),\footnote{
In the context of radio wave propagation in inhomogeneous plasmas,
the analogous problem of linear transformation of modes has been
extensively studied (e.g., Budden~1961; Ginzburg~1970).}
the adiabaticity condition is
\be
\gamma_{\rm res} = (E/E_{\rm ad})^3 \gg 1, \label{eq:adiabat}
\ee
where
\be
E_{\rm ad} = 2.52\,\lb f(B)\tan\thetab\rb^{2/3}|1-\uion|^{2/3}
\! \lp\frac{1\mbox{ cm}}{H_\rho}\rp^{1/3} \!\!\!\! \mbox{keV}
\label{eq:enadiabat}
\ee
and $H_\rho=|dz/d\ln\rho|$ is the density scale height along the ray
(evaluated at $\rho=\densvp$).
For an ionized hydrogen atmosphere,
$H_\rho\simeq 2kT/(\mpr g\cos\theta)=1.65\,T_6/(g_{14}\cos\theta)$~cm,
where $T=10^6T_6$~K is the temperature, $g=10^{14}g_{14}$~cm~s$^{-2}$
is the gravitational acceleration, and $\theta$ is the angle between
the ray and the surface normal.  The probability of a non-adiabatic
``jump'' (e.g., from the $K_+$-curve to the $K_-$-curve) at the
resonance is given approximately by the Landau-Zener formula
$P_{\rm jump}=\exp(-\pi\gamma_{\rm res}/2)$.
Thus, for $E\ga 2E_{\rm ad}$, resonant
conversion between the X-mode and O-mode is essentially complete;
for $E\la 0.5E_{\rm ad}$, a photon will jump across the adiabatic
curves (Fig.~\ref{fig:polark}), and an X-mode (O-mode) photon will
remain an X-mode (O-mode) photon in passing through the resonance.

Note that the adiabatic, vacuum polarization-induced, resonant
mode conversion (from X to O-mode or from O to X-mode) is
an intrinsically coherent phenomenon: the distance over which
the conversion takes place is much smaller than the photon
mean-free path due to absorption or scattering (\laiho).
Such mode conversion is clearly different from the incoherent
mode-switching due to scattering
[see eq.~(\ref{eq:kesji})], which has been included in previous
works (e.g., Pavlov \etal~1995; \"{O}zel~2001; \holai).

As mentioned in Section~\ref{sec:intro}, the formalism of radiative
transfer in strong magnetic fields developed so far is inadequate
for coping with partial mode conversion at the vacuum
resonance.  Thus, in this paper, we shall solve the transport
equations in the two limiting cases: no mode conversion
[when the adiabatic condition given by eq.~(\ref{eq:adiabat}) is never
satisfied] and complete mode conversion (when the adiabatic condition
is always satisfied).  In practice, these correspond to two different
ways of labeling the modes, i.e., X, O-modes with $j=1,2$ and plus
and minus-modes with $j=\pm$ (see Sections~\ref{sec:numcomp} and
\ref{sec:results}).\footnote{Bezchastnov \etal~(1996) apparently
used equation~(\ref{eq:polarkpm}) as the basis of the mode;
therefore they automatically included (complete) mode conversion.
\"{O}zel~(2001) used equation~(\ref{eq:polark}) as the basis and
therefore did not include the mode conversion effect.}

\subsection{Mode Collapse and Breakdown of Faraday Depolarization}
\label{sec:mcp}

Here we comment on the phenomenon of mode collapse
(see also Gnedin \& Pavlov~1974; Pavlov, Shibanov, \& Yakovlev~1980;
Soffel \etal~1983; Bulik \& Pavlov~1996), which appears
under more restrictive conditions than the resonances discussed in
Section~\ref{sec:vpres} [see discussion following eq.~(\ref{eq:polarklo})].
Mode collapse occurs when the two polarization modes become identical,
i.e., $K_1=K_2$ or $K_+=K_-$ [eqs.~(\ref{eq:polark}) and (\ref{eq:polarkpm})],
which requires $\polarb=\pm i\sqrt{\vpr}\approx\pm i$.
Obviously this is only possible when the dissipative terms, i.e.,
those involving $\lambdad=1+i\nu_s/\omega$, in the dielectric
tensor [eq.~(\ref{eq:epsij0})] are retained.  Also, we see from
equation~(\ref{eq:nrefract2}) that the two modes have the same
indices of refraction when mode collapse occurs.

To determine the mode collapse point, we use equation~(\ref{eq:polarb1}),
i.e.,
\be
\polarb = -\frac{\varepsilon'^2-g^2-\varepsilon'\eta'
\lp 1+\vpm/\vpa\rp}{2g\eta'}\frac{\sin^2\thetab}{\cos\thetab}.
 \label{eq:polarb2}
\ee
Note that $\varepsilon', g, \eta'$, and $\polarb$ are considered
complex in this subsection.
The condition $\mbox{Re}(\polarb)=0$ yields the three
critical energies $E^{\rm c}\approx \Evp, \Ebe, \Ebi$, and these
are the resonant energies discussed in Section~\ref{sec:vpres}.
Mode collapse also requires $\mbox{Im}(\polarb)=\pm 1$, so that
$\polarb=\pm i$.  From equation~(\ref{eq:polarb2}), it is easily
shown that
\be
\mbox{Im}(\polarb)=\polarbmcp\gamdampe\frac{\sin^2\thetab}{\cos\thetab},
 \label{eq:polarbim}
\ee
where $\polarbmcp$ depends on the photon energy $E=\hbar\omega$, magnetic
field $B$, and density $\rho$ and
$\gamdampe=\nudamp/\omega=\gamdampce+\gamdampre$ is the dimensionless
damping rate of the electron. For a hydrogen plasma, the collisional
and radiative damping rates are
\ba
\gamdampce & \approx & 9.24\times 10^{-5} \rho_1 T_6^{-1/2} E_1^{-2}
 \lp 1-e^{-E/kT}\rp \\
\gamdampre & = & 9.52\times 10^{-6} E_1, \label{eq:dampe}
\ea
respectively, where we have neglected the Gaunt factor.
The damping rate of the ion is $\gamdampi\approx\gamdampe\me/\mpr$.
Clearly, at a given $E$ ($\approx \Evp, \Ebe, \Ebi$), the condition
$\mbox{Im}(\polarb)=\pm 1$ selects out a critical angle $\thetab=\thetamcp$
at which mode collapse occurs.
For example, for $\rho=1$~g~cm$^{-3}$ and $B=10^{14}$~G,
$|\polarbmcp| \approx 350, 1.3\times 10^6, 0.21$
at $E\approx\Evp, \Ebe, \Ebi$, respectively, and for $B=10^{13}$~G,
$|\polarbmcp| \approx 5.6, 130, 0.63$.\footnote{
At $E\approx\Evp$, one can show that
$\polarbmcp \approx -\uel^{1/2}/2=-\Ebe/(2\Evp)$ in the limit
$\uel\gg 1$ and $\Evp\gg\Ebi$.
}
Since $|\polarbmcp|\gamdampe\ll 1$ at $E\approx\Evp$ and $\Ebi$ for
these parameters, we
see that the mode collapse associated with the vacuum resonance
and the ion cyclotron resonance occur at $\thetab=\thetamcp$
close to 90$^\circ$.

More generally, the modal description of radiative transfer is
valid only in the limit of Faraday depolarization, i.e., when
the condition $|\mbox{Re}(n_+-n_-)|\gg|\mbox{Im}(n_++n_-)|$
is satisfied (Gnedin \& Pavlov~1974).  One can show that away from
the mode collapse points, this is indeed the case.  The breakdown
of Faraday depolarization at $E=\Ebi$ appears to be concentrated
near $\thetab=90^\circ$, but the situation for the vacuum resonance
is more complicated.  For $\uel\gg 1$ (and neglecting the ion effect),
we find
\be
\mbox{Re}(n_+^2-n_-^2) \approx -2\vel\uel^{-1/2}\cos\thetab
 \mbox{Re}\lp\sqrt{1+\polarb^2}\rp \label{eq:mcpreal}
\ee
\be
\mbox{Im}(n_+^2+n_-^2) \approx \gamdampe\vel\lb\sin^2\thetab
 + \uel^{-1}\lp 1+\cos^2\thetab\rp\rb, \label{eq:mcpimag}
\ee
where
\be
\polarb \approx \frac{\uel^{1/2}}{2}\frac{\sin^2\thetab}{\cos\thetab}
 \lp\polarbvp-i\gamdampe\rp \label{eq:polarbmcp}
\ee
and $\polarbvp\approx 1-(\vpq+\vpm)/\vel=1-(E/\Evp)^2$
[see eq.~(\ref{eq:polarbvp})].  Thus, for $|\polarbvp|\ga\gamdampe$
(i.e., for $E$ slightly away from $\Evp$),
Faraday depolarization is satisfied for all $\thetab$.
On the other hand, for $|\polarbvp|\ll\gamdampe$
(i.e., for $E$ extremely close to $\Evp$), the condition
$|\mbox{Re}(n_+-n_-)|\ga|\mbox{Im}(n_++n_-)|$ requires
$(\gamdampe\uel^{1/2}/2)|\sin^2\thetab/\cos\thetab|=|\mbox{Im}(\polarb)|\la 1$;
this condition translates to $\thetab\la\thetamcp$ or
$\thetab\ga\pi-\thetamcp$.
In other words, the breakdown of Faraday depolarization at
the vacuum resonance ($|\polarbvp|\ll\gamdampe$) is restricted
to an angular region of $\thetab$ around $\pi/2$.
To satisfy Faraday depolarization for most $\thetab$ requires
$\gamdampe\uel^{1/2}/2^{3/2}\ll 1$, where $\gamdampe$ and $\uel$ are
evaluated at the vacuum resonance.  For a given magnetic field $B$,
this requirement implies $E\gg\Emcp(B)$: photons with energy
$E$ much greater than $\Emcp(B)$ encounter the vacuum resonance
[at $\densvp(E)$] at which Faraday depolarization is satisfied
for most $\thetab$; conversely, for $E<\Emcp(B)$, Faraday depolarization
breaks down at the vacuum resonance for a significant range of $\thetab$
around $\pi/2$.  Figure~\ref{fig:mcp} shows $\Emcp(B)$ as a function of
$B$.  Note that when the ion effect is included, the simple expressions
given by equations~(\ref{eq:mcpreal})-(\ref{eq:polarbmcp}) are modified,
and we evaluate the complex $n_+^2$ and $n_-^2$ using
equation~(\ref{eq:nrefract2}) and obtain $\Emcp(B)$ numerically
[for $B_{14}\ga 30$, the function $\Emcp(B)$ is multi-valued
because $\mbox{Re}(n_+^2-n_-^2)$ depends non-monotonically on $E$].
For example, at $B=5\times 10^{14}$~G, photons with
$E\la\Emcp\approx 2$~keV encounter the vacuum resonance
between the X-mode and O-mode decoupling layers
[$\rho_O<\densvp(E)<\rho_X^{(\rm nv)}$; see Section~\ref{sec:densd}
and Fig.~\ref{fig:densd}],
and Faraday depolarization breaks down at this resonance for most
$\thetab$.

Currently it is not clear how significant an effect the breakdown
of Faraday depolarization will have on the radiative transfer.
Given that this breakdown of the modal description occurs for such
a narrow range of $E$ [i.e., $\polarbvp\ll\gamdampe\ll 1$; note
that this range is even narrower than the width of the opacity
feature around $\Evp$; see Section~\ref{sec:opacity} and
eq.~(\ref{eq:vpwidth1}) in particular], the effect should be small;
but one cannot be sure.
The above analysis reinforces our statement in Sections~\ref{sec:intro}
and \ref{sec:modeconv} that a rigorous treatment of radiative transfer
near the vacuum resonance, in general, requires solving the transport
equations for the four Stokes parameters.  We plan to study this and
other related issues in a future paper.
We shall ignore the mode collapse phenomenon and the breakdown of
Faraday depolarization in our calculations in the rest of the paper.

\begin{figure}
\centering
\includegraphics[height=8cm]{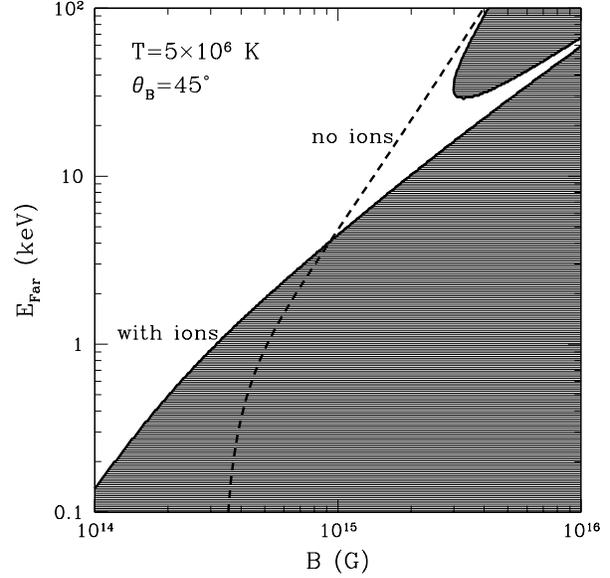}
\caption{
Critical energy $\Emcp(B)$ (at $T=5\times 10^6$~K and $\thetab=45^\circ$)
above which the condition for Faraday depolarization is satisfied at the
vacuum resonance $\rho=\densvp(E)$.
The solid lines indicate $\Emcp$ calculated from
equation~(\ref{eq:nrefract2}) with ions included, while the dashed line
indicates approximate values of $\Emcp$ calculated from
equations~(\ref{eq:mcpreal}) and (\ref{eq:mcpimag}) which neglect ions.
Note that at large magnetic fields ($B\ga 3\times 10^{15}$~G), the
function $\Emcp(B)$ (with ion effects included) is multi-valued.
The unshaded region is where Faraday depolarization is valid, while
the shaded regions are where Faraday depolarization breaks down
(see Section~\ref{sec:mcp} for more details).
\label{fig:mcp}
}
\end{figure}

\setcounter{equation}{0}
\section{Behavior of Opacities and Vacuum Polarization Resonance}
\label{sec:opacity}

The radiative opacities depend on the normal mode polarization
vector through its projection on the rotating frame with the
$z$-axis along $\vechatB$.  The cyclic components of $\vece^j$ are
\ba
|e_\pm^j|^2 & = & \left|\frac{1}{\sqrt{2}}\lp e_x^j\pm ie_y^j\rp\right|^2
 \nonumber \\
 & = & \frac{\lb 1\pm\lp K_j\cos\thetab
 +K_{z,j}\sin\thetab\rp\rb^2}{2(1+K_j^2+K_{z,j}^2)} \label{eq:polarvecpm} \\
|e_z^j|^2 & = &
 \frac{\lp K_j\sin\thetab-K_{z,j}\cos\thetab\rp^2}{1+K_j^2+K_{z,j}^2}
 \label{eq:polarvecz},
\ea
where $K_j$ is given by equation~(\ref{eq:polark}) for $j=1,2$
or equation~(\ref{eq:polarkpm}) for $j=\pm$,
and $K_{z,j}$ is given by equation~(\ref{eq:polarkz}).

The electron scattering opacity from mode $j$ ($=1,2$ for the
X and O-modes or $\pm$ for the plus and minus modes) into mode $i$
is given by (Ventura~1979; Ventura, Nagel, \& M\'{e}sz\'{a}ros~1979)
\be
\kesji = \frac{\nel\sigth}{\rho}\sum_{\alpha=-1}^1
 \frac{\omega^2}{(\omega+\alpha\omegab)^2+\nudamp^2}|e_\alpha^j|^2A_\alpha^i,
 \label{eq:kesji}
\ee
where $\nel$ is the electron number density, $\sigth$ is the
Thomson cross-section, and $A_\alpha^i$ is the angle integral
given by $A_\alpha^i = (3/8\pi)\int d\veck\,|e_\alpha^i|^2$.
The electron scattering opacity from mode $j$ is
\be
\kesj = \frac{\nel\sigth}{\rho}\sum_{\alpha=-1}^1
 \frac{\omega^2}{(\omega+\alpha\omegab)^2+\nudamp^2}|e_\alpha^j|^2 A_\alpha,
 \label{eq:kesj}
\ee
where $A_\alpha=\sum_{i=1}^2 A_\alpha^i$.  In the transverse-mode
approximation ($K_{z,j}=0$), the polarization vector $\vece^j$
satisfies the completeness relation
$\sum_{j=1}^2 |e_\pm^j|^2 = \lp 1+\cos^2\!\thetab\rp/2$
and $\sum_{j=1}^2 |e_z^j|^2 = \sin^2\!\thetab$, and thus $A_\alpha=1$.
For $\omega\ll\omegab$, a suppression factor $(\omega/\omegab)^2$
in the opacity results from the strong confinement of electrons
perpendicular to the magnetic field.  Similar features appear
in the electron free-free absorption opacity
(e.g., Virtamo \& Jauho~1975; Pavlov \& Panov~1976; Nagel \& Ventura~1983)
\be
\kffj = \frac{\alpha_0}{\rho}\sum_{\alpha=-1}^1
 \frac{\omega^2}{(\omega+\alpha\omegab)^2+\nudamp^2}|e_\alpha^j|^2\gff_\alpha,
 \label{eq:keffj}
\ee
where
\ba
\alpha_0 & = & 4\pi^2 Z^2\alphaf^3\frac{\hbar^2 c^2}{\me^2}
 \lp\frac{2\me}{\pi kT}\rp^{1/2} \frac{\nel\nion}{\omega^3}
 \lp 1-e^{-\hbar\omega/kT}\rp \nonumber \\
 & = & \alpha^{\rm ff}_0 \frac{3\sqrt{3}}{4\pi}
 \frac{1}{\gff}, \label{eq:kffmagcoef}
\ea
$\nion$ is the ion number density, and
$\alpha^{\rm ff}_0$ and $\gff$ are the free-free
absorption coefficient and velocity-averaged free-free Gaunt
factor, respectively, in the absence of a magnetic field.
In equation~(\ref{eq:keffj}), $\gff_{\pm 1}=\gff_\perp$ and
$\gff_0=\gff_\parallel$ are the velocity-averaged free-free
Gaunt factors in a magnetic field, which we evaluate using
the expressions given in M\'{e}sz\'{a}ros~(1992)
(see also Nagel~1980).

The ion contribution to the scattering and absorption opacities are
\be
\kisji = \lp\frac{Z^2\me}{A\mpr}\rp^2\!\! \frac{\nion\sigth}{\rho}
 \!\!\sum_{\alpha=-1}^1\!\!\frac{\omega^2}{(\omega-\alpha\omegabi)^2+\nudampi^2}
 |e_\alpha^j|^2A_\alpha^i \label{eq:kisji}
\ee
\be
\kisj = \lp\frac{Z^2\me}{A\mpr}\rp^2\!\!\frac{\nion\sigth}{\rho}
 \!\!\sum_{\alpha=-1}^1\!\!\frac{\omega^2}{(\omega-\alpha\omegabi)^2+\nudampi^2}
 |e_\alpha^j|^2 A_\alpha \label{eq:kisj}
\ee
\be
\kiffj = \frac{1}{Z^3}\!\!\lp\frac{Z^2\me}{A\mpr}\rp^2\!\!\frac{\alpha_0}{\rho}
 \!\!\sum_{\alpha=-1}^1\!\!\frac{\omega^2}{(\omega-\alpha\omegabi)^2+\nudampi^2}
 |e_\alpha^j|^2\gff_\alpha. \label{eq:kiffj}
\ee
Note that the ion cyclotron resonance occurs for $\alpha=+1$, i.e.,
when the electric field of the mode rotates in the same direction
as the ion gyration.
The total scattering and absorption opacities in a fully ionized
medium is then the sum of the electron and ion components, namely
$\kscj = \kesj + \kisj$ and $\kabsj = \kffj + \kiffj$.

In equations~(\ref{eq:kesji})-(\ref{eq:kiffj}), we have included
damping through $\nudamp=\nudampre+\nudampce$
and $\nudampi=\nudampri+\nudampci$, where
$\nudampre=(2e^2/3\me c^3)\omega^2$ and
$\nudampri=(Z^2\me/A\mpr)\nudampre$ are radiative damping rates
and $\nudampce=(\alpha_0\gff_\alpha/\nel\sigth)\nudampre$
and $\nudampci=(\me/A\mpr)\nudampce$ are collisional damping rates
(see Pavlov \etal~1995, and references therein).
For the photon frequencies of interest, $\omega\gg\nudamp,\nudampi$,
and therefore damping is negligible except near resonance.

\begin{figure*}
\centering
\includegraphics[height=10cm]{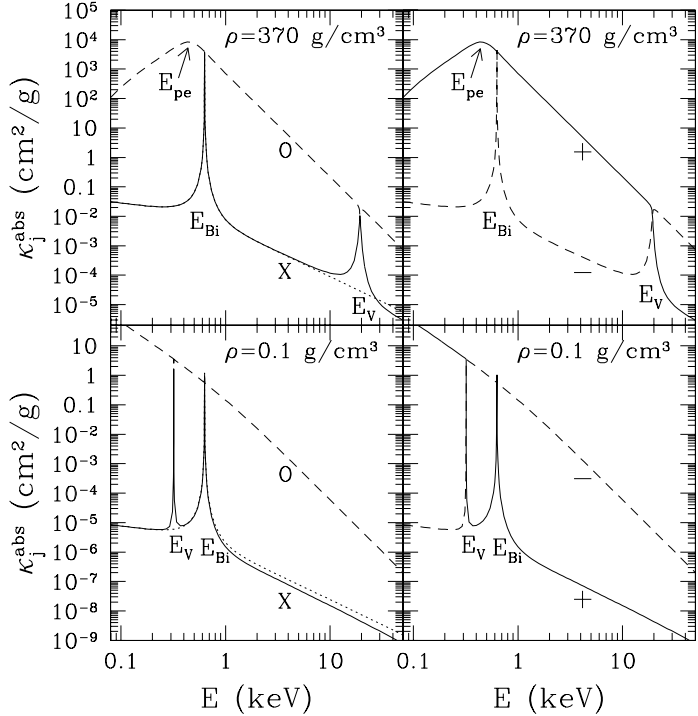}
\caption{Absorption opacity $\kabsj$ as a function of
energy for $B=10^{14}$~G, $\thetab=45^\circ$, and $T=7\times 10^6$~K
at $\rho=370\,\mbox{g cm$^{-3}$}$ (upper panels) and
$\rho=0.1\,\mbox{g cm$^{-3}$}$ (lower panels).
The left panels show the opacities for the X-mode ($j=1$, solid lines)
and the O-mode ($j=2$, dashed lines), while the right panels are for
the plus-mode ($j=+$, solid lines) and the minus-mode ($j=-$, dashed lines).
In the left panels, the dotted lines show the the X-mode opacity
when vacuum polarization is neglected.
Several features are marked in the figure: $\Ebi$ is the ion
cyclotron energy, $\Evp$ is the vacuum resonance energy, and
$\Epe$ is the electron plasma energy.
The variations of the O-mode opacity at $E\lo\Epe$
are due to $K_{z,j}$ [$\propto\omegap^2$; see eq.~(\ref{eq:polarkz})].
\label{fig:ken}
}
\end{figure*}

Figure~\ref{fig:ken} shows examples of the absorption opacity
$\kabsj$ as a function of photon energy for magnetic field
$B=10^{14}$~G and $\thetab=45^\circ$
at densities and a temperature characteristic of NS atmospheres.
For angles $\thetab$ not too close to $0^\circ$ or $180^\circ$
(e.g., $10^\circ\la\thetab\la 170^\circ$), the opacity
$\kabsj(\thetab=45^\circ)$ is indicative of the behavior of
$\kabsj(\veck)$, while $\kabsj(\veck)$ exhibits strong angle dependence
for $\thetab$ near $0^\circ$ or $180^\circ$.
Because of mode conversion due to the vacuum resonance
(see Section~\ref{sec:modeconv}), two sets of photon modes are
depicted in Fig.~\ref{fig:ken}: when mode conversion is neglected
(e.g., an X-mode photon remains in the X-mode when traversing
the vacuum resonance), we use $j=1,2$ for the X and O-modes (left
panels); on the other hand, if we assume that mode conversion
is complete for all energies (e.g., a plus-mode photon remains
in the plus-mode when crossing the resonance), the opacities
are determined with $j=\pm$ (right panels).\footnote{
In plotting the right panels of Fig.~\ref{fig:ken}, we have used
$\polarb=|\polarb_0|\polarbvp$ [see eqs.~(\ref{eq:polarb}) and
(\ref{eq:polarb0})] in evaluating the polarization parameter
$\polarb$ so that the plus or minus-mode is resonant at the
ion cyclotron energy.  If we use $\polarb=\polarb_0\polarbvp$,
then the $+$ and $-$ curves will switch across $\Ebi$ since $\polarb$
changes sign.  However, since $\Ebi$ is independent of $\rho$,
it is easy to show that the radiative transfer is the same for
both cases ($\polarb=|\polarb_0|\polarbvp$ versus
$\polarb=\polarb_0\polarbvp$). \label{foot:ken}
}
Note that since $\Evp$ depends on $\rho$, the left and right
panels of Fig.~\ref{fig:ken} represent genuinely different
opacities in the radiative transfer (see below).

The behavior of the opacities near the ion cyclotron resonance
$\Ebi=0.63\,B_{14}$~keV (for $Z=A=1$) was studied in \holai.
The vacuum resonance feature at $\Evp$ [eq.~(\ref{eq:evp})]
arises from the interplay between the plasma and
vacuum polarization effects discussed in Section~\ref{sec:vpres}
(e.g., M\'{e}sz\'{a}ros \& Ventura~1979; Pavlov \& Shibanov~1979;
Ventura, Nagel, \& M\'{e}sz\'{a}ros~1979).
To understand the behavior of the X and O-mode ($j=1,2$) opacities
near the vacuum resonance, we write the scattering and absorption
opacities as
\ba
\kscj=\frac{\nel\sigth}{\rho}\opbrace_j & , &
\kabsj=\frac{\alpha_0}{\rho}\opbrace_j, \label{eq:opbrace0}
\ea
with
\begin{displaymath}
\opbrace_j \approx \lp\frac{1}{\uel}|e_+^j|^2
 +\frac{1}{\uel}|e_-^j|^2 +|e_z^j|^2\rp + \frac{1}{M'^2}
\end{displaymath}
\be
\times \lb\frac{1}{\lp 1-\sqrt{\uion}\rp^2+\gamdampi^2}|e_+^j|^2
 + \frac{1}{\lp 1+\sqrt{\uion}\rp^2}|e_-^j|^2 + |e_z^j|^2 \rb,
\label{eq:opbrace}
\ee
where $M'=(A\mpr/Z\me)Z^{-1/2}$ for scattering and
$M'=(A\mpr/Z\me)Z^{1/2}$ for absorption, $\gamdampi=\nudampi/\omega$,
and we have set the free-free Gaunt factors to unity in $\kabsj$
and $A_\alpha=1$ in $\kscj$ for simplicity.
Away from the ion cyclotron resonance, the ion contribution to
the opacities is negligible, and
$\opbrace_j\approx(|e_+^j|^2+|e_-^j|^2)/\uel + |e_z^j|^2$.
In the limit of $|\polarb|\gg 1$ (and using the transverse approximation
$K_{z,j}\approx 0$ or $\vel\ll 1$), we find from equations~(\ref{eq:Kjhigh}),
(\ref{eq:polarvecpm}), and (\ref{eq:polarvecz}) that
\be
|e_\pm^1|^2 = \frac{1}{2}\lp 1\mp\frac{1}{\polarb}\cos\thetab\rp , \quad
|e_z^1|^2 = \frac{1}{\lp 2\polarb\rp^2}\sin^2\thetab
 \label{eq:polarvec1approxhigh}
\ee
\begin{displaymath}
|e_\pm^2|^2 = \frac{1}{2}\cos^2\thetab\lp 1\pm\frac{1}{\polarb\cos\thetab}\rp,
\end{displaymath}
\be
|e_z^2|^2 = \lb 1-\frac{1}{\lp 2\polarb\rp^2}\rb\sin^2\thetab.
 \label{eq:polarvec2approxhigh}
\ee
Thus, for the X and O-modes,\footnote{
Note that in this section, we are considering $\thetab$ away from
$0$ and $\pi$; when $\thetab$ is close to $0$ or $\pi$, the dominant
terms in $\opbrace_j$ are different. \label{foot:opbrace}}
\be
\opbrace_1 \approx \frac{1}{\uel}+\frac{1}{4\polarb^2}\sin^2\thetab,
 \quad \opbrace_2 \approx \sin^2\thetab \quad (|\polarb|\gg 1).
 \label{eq:opbracehi}
\ee
On the other hand, in the limit of $|\polarb|\ll 1$, we find
from equation~(\ref{eq:polarklo}) that
\ba
|e_\pm^j|^2 & = & \frac{1}{4}\left\{\lb 1\pm (-1)^j\mbox{sign}(\polarb)
 \cos\thetab\rb^2 \right. \nonumber \\
 & & \left. -\lb(-1)^j|\polarb|+\frac{\vprtwo}{2}\rb
 \sin^2\thetab\right\} \\
|e_z^j|^2 & = & \frac{1}{2}\lb 1+\lp -1\rp^j|\polarb|+\frac{\vprtwo}{2}\rb
 \sin^2\thetab \label{eq:polarvecapproxlow}.
\ea
Thus we have
\be
 \opbrace_j \approx \frac{1}{2}\sin^2\thetab\lb 1+(-1)^j|\polarb|\rb
\qquad (|\polarb|\ll 1).  \label{eq:opbracelo}
\ee
Clearly the X-mode ($j=1$) opacity is significantly enhanced near the
vacuum resonance ($\polarb\sim 0$), while the O-mode opacity
is only reduced by a factor of two from its usual value.
At the resonance, the two modes become circularly polarized and
possess the same opacities (see left panels of Fig.~\ref{fig:ken}).

Equations~(\ref{eq:opbracehi}) and (\ref{eq:opbracelo}) determine
the ``line shape'' of the vacuum resonance feature.
For $x\equiv(E-\Evp)/\Evp\ll 1$,
equations~(\ref{eq:polarb1})-(\ref{eq:polarbvp0}) imply
$|\polarbvp|\approx 2|x|$ and
\be
|\polarb|\approx\uel^{1/2}|1-\uion||x|\frac{\sin^2\thetab}{\cos\thetab}
=\frac{|x|}{x_{\rm V}} \label{eq:vpbetares}
\ee
with
\be
x_{\rm V}\equiv\frac{\cos\thetab}{\uel^{1/2}|1-\uion|\sin^2\thetab}.
\label{eq:vpwidth1}
\ee
Thus
\be
\opbrace_X=\opbrace_1\approx
 \left\{ \begin{array}{ll}
 \frac{1}{2}\sin^2\thetab\lp 1-|x|/x_{\rm V}\rp
 & |x|\ll x_{\rm V} \\
 \uel^{-1}+(x_{\rm V}/2x)^2\sin^2\thetab
 & x_{\rm V} \ll |x| \ll 1
 \end{array} \right. . \label{eq:opbracevp}
\ee
This should be compared to the value when there is no vacuum
polarization (nv),
\be
\opbrace_X^{(\rm nv)} \approx \uel^{-1} +x_{\rm V}^2\sin^2\thetab
 \sim \uel^{-1}.  \label{eq:opbracenvp}
\ee
The strength of the vacuum resonance feature
can be measured by the dimensionless ``equivalent width''
\ba
\Gammavp & \equiv & \frac{1}{\kappa_X^{(\rm nv)}\Evp}\int_{\rm res}\kappa_X dE
 \nonumber \\
 & = & \frac{1}{\opbrace_X^{(\rm nv)}\Evp}\int_{\rm res}\opbrace_X dE
 \sim \uel\int_{-1/2}^{1/2}\opbrace_X dx, \label{eq:vpwidth2}
\ea
where the integral is over the region in which
$\opbrace_X/\opbrace_X^{(\rm nv)}>1$,
i.e., $\Evp/2\la E\la 3\Evp/2$ or $-1/2\la x\la 1/2$
(see Fig.~\ref{fig:vpline}).
Using equation~(\ref{eq:opbracevp}) and for
$\sin^2\thetab\sim\cos\thetab\sim 1$, we find
\be
\Gammavp \sim \frac{\uel^{1/2}}{|1-\uion|} \approx \frac{10^3}{|1-\uion|}
 \lp\frac{\mbox{1 keV}}{E}\rp B_{14}.  \label{eq:vpwidth3}
\ee
It is important to note that the region $|x|\la(\mbox{a few }x_{\rm V})\ll 1$
contributes most to the integral in equation~(\ref{eq:vpwidth2}).
This is illustrated by the shaded region in Figure~\ref{fig:vpline},
which shows a schematic picture of the vacuum
resonance feature; note that from equation~(\ref{eq:opbracevp}),
the continuum is $\uel^{-1}$ lower than the peak.
This region must be resolved in numerical calculations of
atmosphere spectra (see Section~\ref{sec:numcomp}).
Also note that at low densities where $\Evp$ lies below the ion
cyclotron energy so that $\uion > 1$, the vacuum resonant feature
is narrower [see eq.~(\ref{eq:vpwidth1}); also compare the two
left panels in Fig.~\ref{fig:ken}].

\begin{figure}
\centering
\includegraphics[height=8cm]{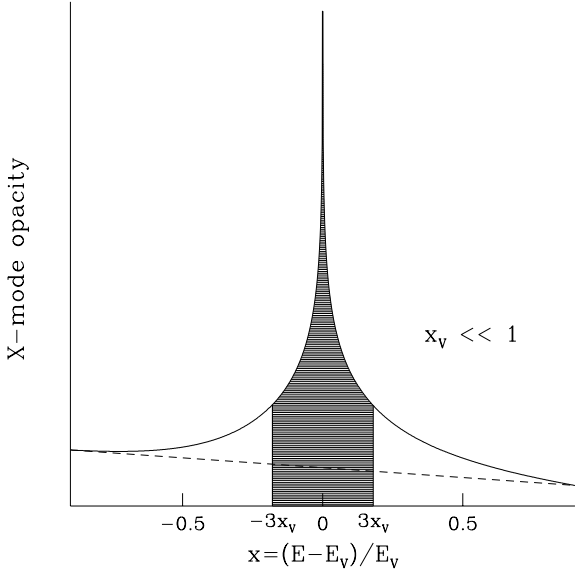}
\caption{
A schematic picture of the X-mode opacity near the vacuum resonance
energy $\Evp$.  The dashed line represents the opacity when vacuum
polarization is neglected.  The shaded region,
$-3x_{\rm V}\lo x\lo 3x_{\rm V}$ [see eq.~(\ref{eq:vpwidth1});
note that $x_{\rm V}\ll 1$], provides the dominant contribution
to the equivalent width of the vacuum resonance feature
[eqs.~(\ref{eq:vpwidth2}) and (\ref{eq:vpwidth3})], although a
much broader region, $-0.5\lo x\lo 0.5$, is affected by vacuum
polarization.  The opacity at the resonance ($E=\Evp$) is
$\sim\uel$ times larger than the continuum.
\label{fig:vpline}
}
\end{figure}

When mode conversion is assumed to be complete, the plus and
minus-mode opacities $\kappa_\pm$ do not exhibit a line feature
but rather show a step function-like change at the vacuum
resonance energy $\Evp$ (see the right panels in Fig.~\ref{fig:ken}).

\setcounter{equation}{0}
\section{Qualitative Discussion of the Vacuum Polarization Effect
on Atmosphere Emission}
\label{sec:densd}

In order to understand qualitatively the effects of vacuum polarization
on the radiation spectra from magnetar atmospheres, we estimated in
\laiho\ the location of the decoupling layer at which the optical
depth is of order unity for photons of different energies and
polarization modes.
Here we conduct a somewhat more accurate calculation by including
the energy dependence of the Gaunt factor and show
results for different magnetic field strengths.
For simplicity, we consider fully ionized hydrogen atmospheres,
adopt the transverse approximation ($K_{z,j}=0$), and present
results for only $\thetab=45^\circ$.
We also treat the temperature as a constant, which is a good
approximation when estimating the decoupling density since $T$
varies at most by a factor of a few while $\rho$ varies by many
orders of magnitude above the decoupling layer.
For $g=(GM/R^2)(1-2GM/Rc^2)^{-1/2}=2.4\times 10^{14}$~cm~s$^{-2}$
(corresponding to a $M=1.4\,M_\odot$, $R=10$~km NS),
hydrostatic equilibrium and the ideal gas equation of state,
$P=2\rho kT/\mpr$, give the density $\rho$ at column density
$y$ (in g cm$^{-2}$) as $\rho=1.45\,T_6^{-1}y$~g cm$^{-3}$,
where $T_6=T/(10^6\mbox{ K})$.
The optical depth $\tau_\nu$ ($=\int\kappa_\nu dy$) at $\rho$
is then given by
$\tau_\nu=0.69\,T_6\int_0^\rho\kappa_\nu\,d\rho'$.
Note that we use here the effective opacity defined by
$\kappa_\nu=[\kabs_\nu(\kabs_\nu+\ksc_\nu)]^{1/2}$
(see Rybicki \& Lightman~1979).
The decoupling density $\rho$ can then solved for by setting the
optical depth $\tau_\nu=1$.
Numerical results of the decoupling density are shown in
Figure~\ref{fig:densd} for $B=10^{14}$ and $5\times 10^{14}$~G.

\begin{figure}
\centering
\includegraphics[height=10cm,width=8.5cm]{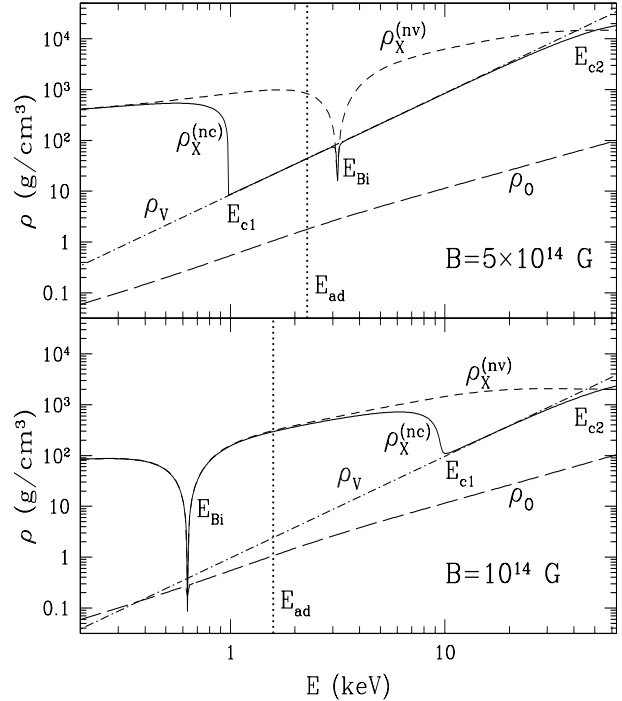}
\caption{
The density at which photons of different polarization modes
decouple from the matter as a function of photon energy for
$T=5\times 10^6$~K, $\thetab=45^\circ$,
and $B=10^{14}$~G (lower panel) and $B=5\times 10^{14}$~G (upper panel).
The solid lines show the X-mode decoupling density $\rho_X^{(\rm nc)}$
when vacuum polarization is included but mode conversion is neglected,
the short-dashed lines are for $\rho_X^{(\rm nv)}$ with no
vacuum polarization, the long-dashed lines are for the
O-mode $\rho_O$ (note that $\rho_O$ is unaffected by the magnetic
field and vacuum polarization effects), and
the dot-dashed lines show the vacuum resonance density $\densvp$
[eq.~(\ref{eq:densvp})].
The vertical dotted lines denote the critical energy $E_{\rm ad}$
for adiabatic mode conversion [eq.~(\ref{eq:enadiabat})].
In general, vacuum polarization (including the mode conversion
effect) reduces the decoupling density of X-mode photons with
energies between $\min(E_{\rm c1},E_{\rm ad})$ and $E_{\rm c2}$,
where $E_{\rm c1}$ and $E_{\rm c2}$ are given approximately by
equations~(\ref{eq:ec1}) and (\ref{eq:ec2}), respectively.
The ion cyclotron resonance $\Ebi=0.63\,B_{14}$ is clearly evident,
though it can be suppressed when vacuum polarization is taken into
account, as in the case of $B=5\times 10^{14}$~G.
\label{fig:densd}
}
\end{figure}

We can obtain analytic expressions to understand the various
features in Fig.~\ref{fig:densd} by setting the Gaunt factor to unity
and by ignoring scattering, which is only important for $E\ga 10$~keV,
even though the results shown in Fig.~\ref{fig:densd} do account for
the actual variation of the Gaunt factor and use the effective
opacity which includes scattering.
First, let us ignore vacuum polarization and proton effects.
The absorption opacity can then be written as
$\kabsj\approx\kappa_0\opbrace_j$, where
$\kappa_0\approx 9.3\,\rho_1T_6^{-1/2}E_1^{-3}S$~cm$^2$g$^{-1}$
[with $S=1-e^{-E/kT}$, $\rho_1=\rho/(1\mbox{ g cm$^{-3}$})$,
and $E_1=E/(1\mbox{ keV})$] is the zero-field opacity and
$\opbrace_O\sim 1$ and $\opbrace_X\sim\uel^{-1}$
[see eqs.~(\ref{eq:opbrace0}) and (\ref{eq:opbracehi})].
The decoupling densities of the two modes [with no vacuum polarization
(nv)] are then approximately
\ba
\rho_O^{(\rm nv)} & \approx & 0.56\,T_6^{-1/4}E_1^{3/2}S^{-1/2}
 \mbox{ g cm$^{-3}$} \\
\rho_X^{(\rm nv)} & \approx & \uel^{1/2}\rho_O^{(\rm nv)}.
\ea
Thus the X-mode photons emerge from deeper in the atmosphere and
are the main carriers of the X-ray flux.

Next, we include vacuum polarization and protons but neglect resonant
mode conversion.  For the O-mode, the opacity is largely unaffected
by the magnetic field and vacuum polarization effect.  Thus the
decoupling density $\rho_O$ is still $\rho_O^{(\rm nv)}$.
For the X-mode, when $\rho_X^{(\rm nv)}\ga\densvp$ or
\be
E\la E_{\rm c2} \approx 77\,T_6^{-1/6}B_{14}^{-2/3}f(B)^{4/3}\mbox{ keV},
 \label{eq:ec2}
\ee
the photons created at $\rho_X^{(\rm nv)}$ will encounter the
vacuum resonance, near which the X-mode opacity is greatly enhanced;
thus the decoupling density [with no mode conversion (nc)]
$\rho_X^{(\rm nc)}$ will be smaller than $\rho_X^{(\rm nv)}$.
For $E\ga E_{\rm c2}$, $\rho_X^{(\rm nc)}$ is close to $\rho_X^{(\rm nv)}$,
with the small difference due to the opacity being slightly modified
by the vacuum effect even away from the resonance (see the left panels of
Fig.~\ref{fig:ken}).
To evaluate the X-mode optical depth $\Delta\tau$ across the resonance
region (e.g., $0.9 < \rho/\densvp < 1.1$) we note that
$\polarb\sim\uel^{1/2}x$ [eq.~(\ref{eq:vpbetares})]
for $|x|\equiv|\Delta\rho/\densvp|\la 0.1$.
Using equations~(\ref{eq:opbracehi}) and (\ref{eq:opbracelo}),
we find $\opbrace_X\sim 1/(\uel x^2)$ for $\uel^{-1/2}\ll|x|\la 0.1$
and $\opbrace_X\sim 1$ for $|x|\ll\uel^{-1/2}$
[compare with eq.~(\ref{eq:opbracevp})].
We then have $\Delta\tau\sim 0.7\,\kappa_0T_6\densvp\uel^{-1/2}
=6.4\,T_6^{1/2}\densvp^2E_1^{-3}\uel^{-1/2}S$,
which is a factor $\sim\uel^{1/2}$ larger than the optical depth
of the non-resonant region ($\rho\la 0.9\,\densvp$).
Thus at energies below $E_{\rm c2}$, $\rho_X^{(\rm nc)}$
closely follows $\densvp$ until $E$ drops below another
critical energy $E_{\rm c1}$, which is set by $\Delta\tau\approx 1$ or
\be
E_{\rm c1} \sim 10\,T_6^{-1/4}B_{14}^{-3/2}f(B)^2\mbox{ keV}.  \label{eq:ec1}
\ee
Below $E_{\rm c1}$, $\rho_X^{(\rm nc)}$ returns to approximately
$\rho_X^{(\rm nv)}$ (see Fig.~\ref{fig:densd}).

Now consider the effect of mode conversion at the vacuum resonance.
For $E>1.3\,E_{\rm ad}$ [see eq.~(\ref{eq:enadiabat})], adiabatic mode
conversion is nearly complete ($P_{\rm jump} < 3\%$).  The O-mode
photons traveling from high densities through $\densvp$ are
converted to X-mode photons, which then freely stream out of the
atmosphere.  Thus, for $1.3 E_{\rm ad}\la E\la E_{\rm c2}$, the
effective decoupling density for the X-mode is $\rho_X=\densvp$.
For $E\ga E_{\rm c2}$, the vacuum resonance occurs inside both
the X and O-mode decoupling layers, and therefore
$\rho_X\approx\rho_X^{(\rm nv)}$.  For $E\ll E_{\rm ad}$, mode
conversion is ineffective; therefore $\rho_X=\rho_X^{(\rm nc)}$.
Around $E_{\rm ad}$, the X-mode photons are emitted from both
$\rho_X^{(\rm nc)}$ (with probability $P_{\rm jump}$) and
$\densvp$ [with probability $(1-P_{\rm jump})$].
Also note that $\rho_X^{(\rm nc)}\approx\densvp$ for $E\ga E_{\rm c1}$.

Figure~\ref{fig:densd} and the above analysis show that vacuum
polarization reduces the decoupling density for photons with
energies between $\min(E_{\rm c1},E_{\rm ad})$ and $E_{\rm c2}$,
i.e., these photons decouple in shallower, lower temperature
regions of the atmosphere.  This gives rise to a rather broad
depression feature in the emergent radiation between
$\min(E_{\rm c1},E_{\rm ad})$ and $E_{\rm c2}$ (see also \laiho).
In other words, even though the vacuum resonance opacity feature is
very narrow, it produces a broad depression because of its density
dependence and the large density gradient present in NS atmospheres.
Clearly, to quantify the depth of the depression, one must solve
for the temperature profile of the atmosphere self-consistently
(see Sections~\ref{sec:numcomp} and \ref{sec:results}).

Finally, the sharp absorption feature in Fig.~\ref{fig:densd} is due to
the ion cyclotron resonance at $\Ebi=0.63\,B_{14}$~keV.
Previous atmosphere models which neglect vacuum polarization
yield relatively large equivalent widths for the proton
cyclotron line (\holai; Zane \etal~2001).
However, we see from Fig.~\ref{fig:densd} that when $\Ebi$ lies
within the depression trough, i.e.,
$E_{\rm c1}\la\Ebi\la E_{\rm c2}$ or
$3\times 10^{14}$~G~$\la Bf(B)^{-4/5}T_6^{1/10}\la 2\times 10^{15}$~G,
the continuum flux around $\Ebi$ is greatly reduced,\footnote{
If we use $E_{\rm ad}\la \Ebi$, then the first inequality
becomes $4\times 10^{14}$~G~$\la Bf(B)^{-2/3}(\tan\thetab)^{-2/3}$.
Note that the range of
magnetic fields must be considered very approximate since
$E_{\rm c1}$, $E_{\rm ad}$, and $E_{\rm c2}$ depend on the direction
of photon propagation, and redistribution of photon spectral flux
occurs in real atmospheres.  For example, even for $B=10^{14}$~G, the
width of the ion cyclotron line is reduced by vacuum polarization
(see Section~\ref{sec:spectra}).
}
and the equivalent width of the ion cyclotron line becomes much
narrower.
This suppression of the ion cyclotron line by vacuum polarization
makes the line more difficult to detect (see Section~\ref{sec:spectra}).

In Appendix~\ref{sec:toy}, we present a toy atmosphere model that
mimics the vacuum resonance effect discussed in this section.

\section{Numerical Method for Treating Vacuum Polarization}
\label{sec:numcomp}

We solve the full, angle-dependent radiative transfer equations
for the two coupled photon modes in order to construct self-consistent
NS atmosphere models.
Because the method for rigorously treating partial mode conversion
has not been developed, we consider two limiting cases:
no mode conversion and complete
mode conversion at all energies (see Section~\ref{sec:modeconv}).
In our models, the
temperature corrections $\Delta T(\tau)$ at each Thomson depth $\tau$
are applied iteratively until $\Delta T(\tau)/T(\tau)\la 1\%$,
deviations from radiative equilibrium are $\la 1\%$, and deviations
from constant flux are $\la 2\%$
(see \holai\ for details of our numerical method).
Here we discuss the numerical difficulties involved when attempting
to include the effects of vacuum polarization in the atmosphere models.

\subsection{No Mode Conversion: Grid Resolution}
\label{sec:grid}

As shown in Section~\ref{sec:opacity}, vacuum polarization induces
a narrow, density-dependent resonance feature in the X-mode opacities
(see Figs.~\ref{fig:ken} and \ref{fig:vpline}).
The ``equivalent width'' of the vacuum resonance feature at $\Evp(\rho)$
is dominated by the narrow energy range
$\Delta E/\Evp=|E-\Evp|/\Evp\la 3\,x_{\rm V}\sim\uel^{-1/2}
\sim 10^{-3}E_1/B_{14}$, where $E_1=E/(\mbox{1 keV})$.
In a real atmosphere, density and energy gradients
are smooth so that such a narrow and density-dependent feature
is resolved and accounted for completely.
Numerical atmosphere models, however, necessarily
require finite grids of discrete depth (density), energy, and angle.
The difficulty with a finite grid resolution is illustrated in
Figure~\ref{fig:opdepth}.
Consider a photon with energy $E$ that happens to lie very close
[$|E-\Evp(\rho_i)|/\Evp(\rho_i)\la x_{\rm V}$] to the vacuum
resonance energy $\Evp(\rho_i)$ of one of the density grid points
$\rho_i$; such a photon will encounter a region of strongly enhanced
opacity and will therefore decouple from the matter at a shallow depth.
On the other hand, a photon with energy $E'$, which lies outside
the resonance regions of all density grid points will not
encounter any enhanced opacity regions and will decouple deeper
in the atmosphere.  Therefore, if we choose a general energy
grid that is not tied to the density grid, the computed radiation
spectrum will exhibit narrow absorption lines at energies which
happen to lie close to $\Evp$ at one of the density grid points.
Such narrow lines are numerical artifacts of the finite grid
resolution since, as discussed in Section~\ref{sec:densd},
vacuum polarization is expected to only produce a broad depression
in the emission spectrum.

\begin{figure}
\centering
\includegraphics[height=8cm]{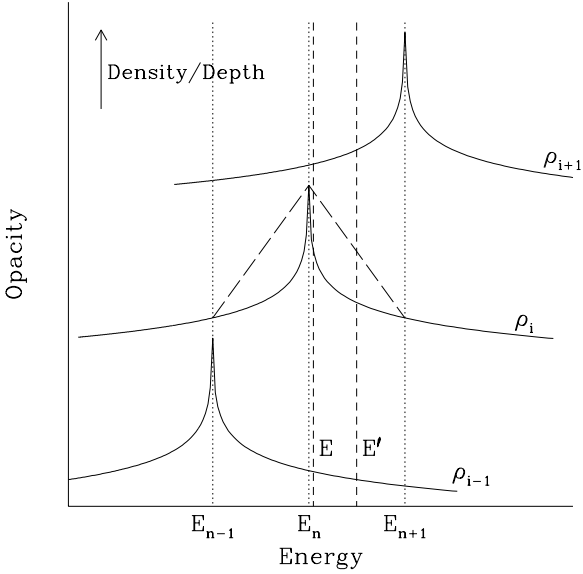}
\caption{
A schematic picture of the radiative opacity as a function of
energy at three neighboring density grid points
$\rho_{i-1}$, $\rho_i$, $\rho_{i+1}$.
Density and depth increase upwards, and the vacuum resonance feature
occurs at a higher energy in deeper levels
[see eqs.~(\ref{eq:evp0}) or (\ref{eq:evp})].
The opacity at a given depth has been offset for clarity.
In the ``equal grid'' method, the energy grid points
$E_{n-1}$, $E_n$, $E_{n+1}$ are chosen to be equal to the vacuum
resonance energy at one of the density grid points, i.e., $E_n=\Evp(\rho_i)$.
The long-dashed line for $\rho_i$ indicates the
apparent opacity obtained by interpolating energy grid points $E_{n-1}$,
$E_n$, and $E_{n+1}$ (see Section~\ref{sec:grid}).
The energy $E$ lies close to one of the resonance energies, while
$E'$ lies outside any resonance region.
\label{fig:opdepth}
}
\end{figure}

Clearly, to correctly account for the density-dependent vacuum
resonance, one must have a depth grid that is sufficiently dense
so that the opacity features at neighboring grid points overlap
appreciably, i.e.,
$[\Evp(\rho_{i+1})-\Evp(\rho_i)]/\Evp(\rho_i)\la x_{\rm V}$
or $(\rho_{i+1}-\rho_i)/\rho_i\la\uel^{-1/2}\sim 10^{-3}E_1B_{14}^{-1}$.
Such a dense depth grid is beyond the capabilities of the current
generation of atmosphere models.

One method to avoid artificial ``absorption lines'' in the computed
spectrum is to adopt an ``equal grid'' method: we use an equal number of
depth (density) and energy grid points with every energy grid point
being placed at $E_n=\Evp(\rho_i)$ (see Fig.~\ref{fig:opdepth}).
The advantage of this method is that it guarantees a smooth spectrum.
However, this method has problems as well.  In order to
determine the atmosphere structure through temperature corrections
(see \holai, and references therein), it is necessary to compute
mean opacities, which involve integrations over energy, at
a given depth;  using the ``equal grid'' method leads to overestimates
of the strength of the opacity feature 
(see Fig.~\ref{fig:opdepth}:
the area under the long-dashed line overestimates the area under
the solid line, which represents the true opacity).
Nevertheless, by increasing the number of depth grid points,
the ``equal grid''
method will yield atmosphere models increasingly close to reality
(see Section~\ref{sec:results}).

An alternative method to produce smooth spectra involves
``saturating'' the opacities near the vacuum resonance (\"{O}zel~2001).
We shall comment on this method in Section~\ref{sec:saturation}.

\subsection{Complete Mode Conversion} \label{sec:numcompmc}

If we assume the adiabatic condition given by equation~(\ref{eq:adiabat})
is satisfied for all energies, then a plus-mode (minus-mode) photon
will remain in the plus-mode (minus-mode) as it traverses the
vacuum resonance (see Section~\ref{sec:modeconv}).
In this ``complete mode conversion'' limit, we simply solve
the coupled radiative transfer equations for the plus and minus-mode
photons rather than the X and O-mode photons.  The numerical
procedure is analogous to the one outlined in \holai.
The modes are calculated using equation~(\ref{eq:polarkpm}) for
the polarization parameter rather than equation~(\ref{eq:polark}).
In this case, the vacuum resonance manifests as a step function-like
feature in the opacities, as discussed in Section~\ref{sec:opacity}
and shown in the right panels of Fig.~\ref{fig:ken}, and there is
no numerical difficulty in handling such an opacity feature.

\setcounter{equation}{0}
\section{Numerical Results} \label{sec:results}

\subsection{Atmosphere Structure} \label{sec:atmstructure}

The temperature profiles for fully ionized hydrogen atmospheres
with $\Teff=5\times 10^6$~K and magnetic field $B=5\times 10^{14}$~G
oriented perpendicular to the surface (i.e., the angle between
the field and the surface normal is $\Thetab=0$)
are plotted in Figure~\ref{fig:tempgrid}.
The curves marked ``nc\#'' are models which include vacuum polarization
but no mode conversion and using the equal grid
method described in Section~\ref{sec:grid}, with the ``\#'' indicating
the number of grid points per decade in Thomson depth $\tau$.
When vacuum polarization is neglected (nv), the temperature
profiles show a plateau at $\tau\sim 1-100$ (such a feature was
already noted in \holai\footnote{
The ``no vacuum polarization'' models presented here differ
slightly from those in \holai\ because of a numerical error
in our calculation of the Gaunt factors in \holai.  This error
has been corrected in the present paper.
}
and in \"{O}zel~2001\footnote{
\"{O}zel's models include vacuum polarization; see
Section~\ref{sec:saturation}.
}
).  This arises
because the X-mode photons decouple at $\tau\sim\mbox{a few}\times 10^2$
and thus have very little energy exchange with the matter,
while for $\tau\ga 1$, the O-mode photons are ineffective at
regulating the temperature because of their small flux.
We see from Fig.~\ref{fig:tempgrid} that vacuum polarization
tends to diminish this plateau in the temperature profile
since vacuum polarization causes the X and O-mode opacities
to be similar to each other (see Fig.~\ref{fig:ken}) and the
decoupling densities to be closer (see Fig.~\ref{fig:densd}).
We also see that vacuum polarization tends to increase the
temperature at most depths (and the general temperature gradient)
relative to the case without vacuum polarization.
In the deepest layers where both modes are diffusing, a rise
in the X-mode opacity due to the vacuum resonance feature reduces
the bandwidth over which flux can be transported, and a larger
temperature gradient is required to maintain a constant total
flux; these layers are therefore backwarmed to higher temperatures
(see Mihalas~1978 for a general discussion of backwarming).
On the other hand, in the plateau $\tau\sim 1-100$,
the free streaming X-mode photons, generated from deeper, hotter
layers, encounter a sharp rise in opacity and are absorbed, which
causes the temperature to increase.

\begin{figure}
\centering
\includegraphics[height=8cm]{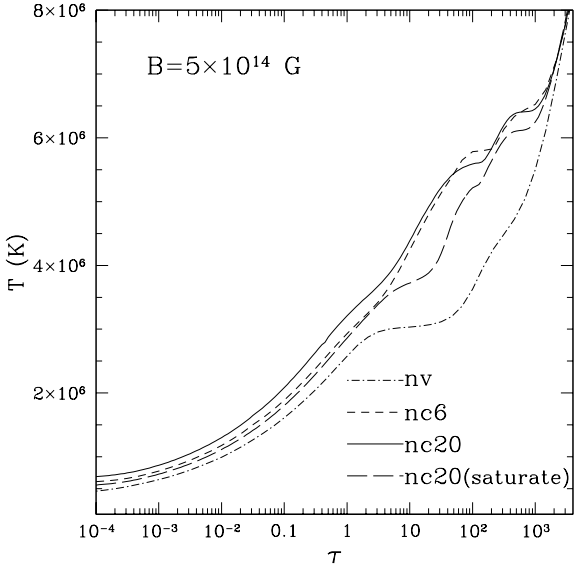}
\caption{
Temperature $T$ as a function of Thomson depth $\tau$
for fully ionized hydrogen atmospheres with
$B=5\times 10^{14}$~G, $\Thetab=0$, and $\Teff=5\times 10^6$~K.
The curves marked ``nc\#'' are models which include vacuum polarization
but no mode conversion
using the ``equal grid'' method described in Section~\ref{sec:grid},
with ``\#'' indicating the number of grid points per decade in $\tau$.
The long-dashed line is for the same model as ``nc20'' but restricting
$\polarbvp\ge 10^{-2}$ (see Section~\ref{sec:saturation}), while
the dot-dashed line is for the model with no vacuum polarization (nv).
\label{fig:tempgrid}
}
\end{figure}

\begin{figure}
\centering
\includegraphics[height=10cm,width=8.5cm]{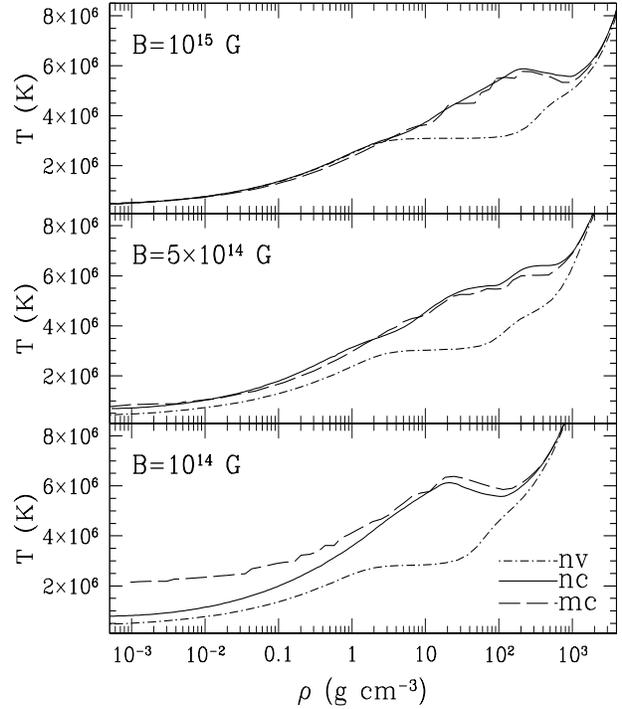}
\caption{
Temperature $T$ as a function of density $\rho$
for fully ionized hydrogen atmospheres with $\Teff=5\times 10^6$~K,
$\Thetab=0$, and
$B=10^{14}$~G (lower panel), $B=5\times 10^{14}$~G (middle panel),
$B=10^{15}$~G (upper panel).
The solid lines are for models with vacuum polarization
but no mode conversion (nc),
the dashed lines are for models which assume complete mode
conversion (mc), and the dot-dashed lines are for models
with no vacuum polarization (nv).
\label{fig:temp}
}
\end{figure}

Figure~\ref{fig:temp} compares the temperature versus density
profiles for various magnetic fields and vacuum polarization
effects.  For the models which include vacuum polarization
but neglect mode conversion (nc), we use twenty grid points per
decade in $\tau$.
Compared to models without vacuum polarization (nv),
the models with complete mode conversion (mc)
show higher temperatures in the deep layers:
although the O-mode photons at a particular depth are
converting into X-mode photons and carrying away heat, X-mode photons
from deeper, hotter layers are converting into O-mode photons and
depositing more energy.
We also note that larger magnetic fields cause the vacuum
resonance to occur at a lower energy for a given density or at
a higher density for a
given energy [see eqs.~(\ref{eq:evp0}), (\ref{eq:evp}), and (\ref{eq:densvp})]
and thus shift the region that is heated by vacuum polarization
effects to deeper layers.

In Appendix~\ref{sec:toy}, we construct toy atmosphere models with
opacities which mimic the vacuum polarization effect.
The temperature profiles of the toy models are qualitatively
similar to the results discussed in this section.

\subsection{Spectra} \label{sec:spectra}

Figure~\ref{fig:spectrum5hgrid} shows the spectra of fully ionized
hydrogen atmospheres with $B=5\times 10^{14}$~G, $\Thetab=0$
(magnetic field perpendicular to stellar surface), and $\Teff=5\times 10^6$~K.
Plotted alongside is the blackbody spectrum at $T=5\times 10^6$~K.
This figure illustrates the effect of grid resolution for models
which include vacuum polarization but no mode conversion.
As discussed in Section~\ref{sec:grid}, a lower grid resolution
tends to overestimate the strength of the vacuum resonance feature,
which leads to the stronger depression at $\sim 0.2-2$~keV.
In the following, we shall use the ``nc20'' model when we show
results which include vacuum polarization but no mode conversion
and adopt the abbreviation ``nc''.

\begin{figure}
\centering
\includegraphics[height=8cm]{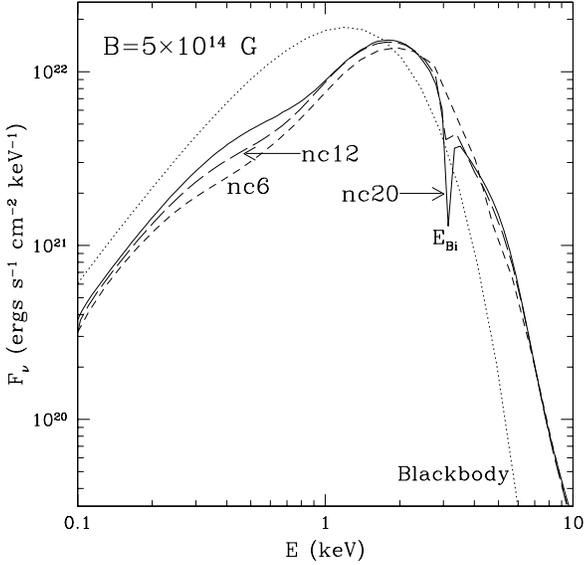}
\caption{
Spectra of fully ionized hydrogen atmospheres with
$B=5\times 10^{14}$~G, $\Thetab=0$, and $\Teff=5\times 10^6$~K.
The curves marked ``nc\#'' are models which include vacuum polarization
but no mode conversion
using the ``equal grid'' method described in Section~\ref{sec:grid},
with ``\#'' indicating the number of grid points per decade in $\tau$.
The corresponding temperature profiles are shown in Fig.~\ref{fig:tempgrid}.
The dotted line is for a blackbody with $T = 5\times 10^6$~K.
Note that the absence of the ion cyclotron feature at $\Ebi$ for
model nc6 is an artifact of the low grid resolution.
\label{fig:spectrum5hgrid}
}
\end{figure}

\begin{figure}
\centering
\includegraphics[height=8cm]{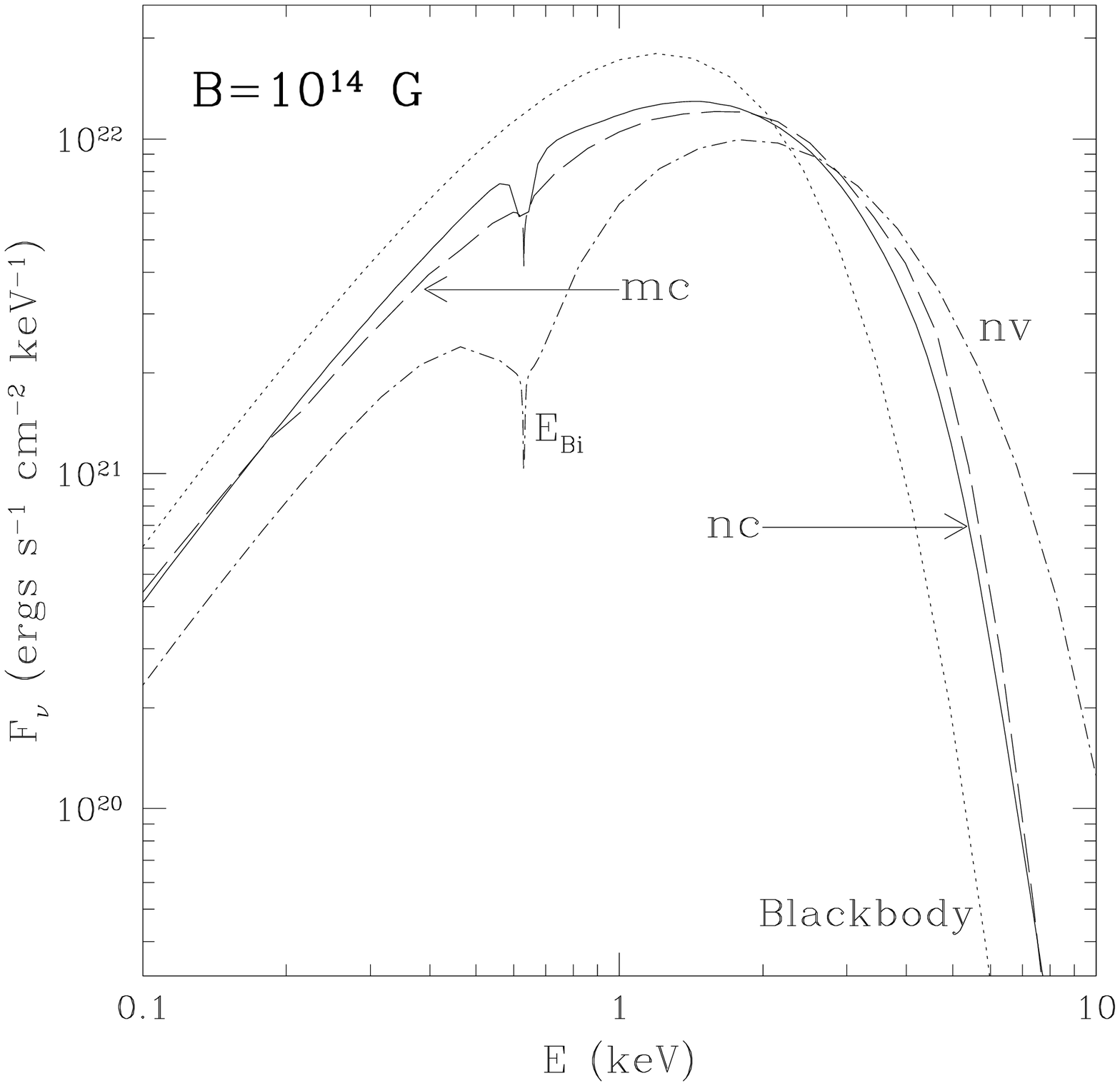}
\caption{
Spectra of fully ionized hydrogen atmospheres with
$B=10^{14}$~G, $\Thetab=0$, and $\Teff=5\times 10^6$~K.
The solid line is for an atmosphere with vacuum polarization
but no mode conversion (nc),
the dashed line is for an atmosphere with complete mode conversion (mc),
the dot-dashed line is for an atmosphere with no vacuum polarization (nv),
and the dotted line is for a blackbody with $T = 5\times 10^6$~K.
The $\Ebi=0.63$~keV ion cyclotron feature (from $\sim 0.4$ to $\sim 1$~keV)
has an equivalent width $\sim 140$~eV when vacuum polarization is not
included (model nv) but is only $\sim 20$~eV in the no mode conversion case
(model nc) and $\sim 6$~eV in the complete mode conversion case (model mc).
\label{fig:spectrum5h14}
}
\end{figure}

\begin{figure}
\centering
\includegraphics[height=8cm]{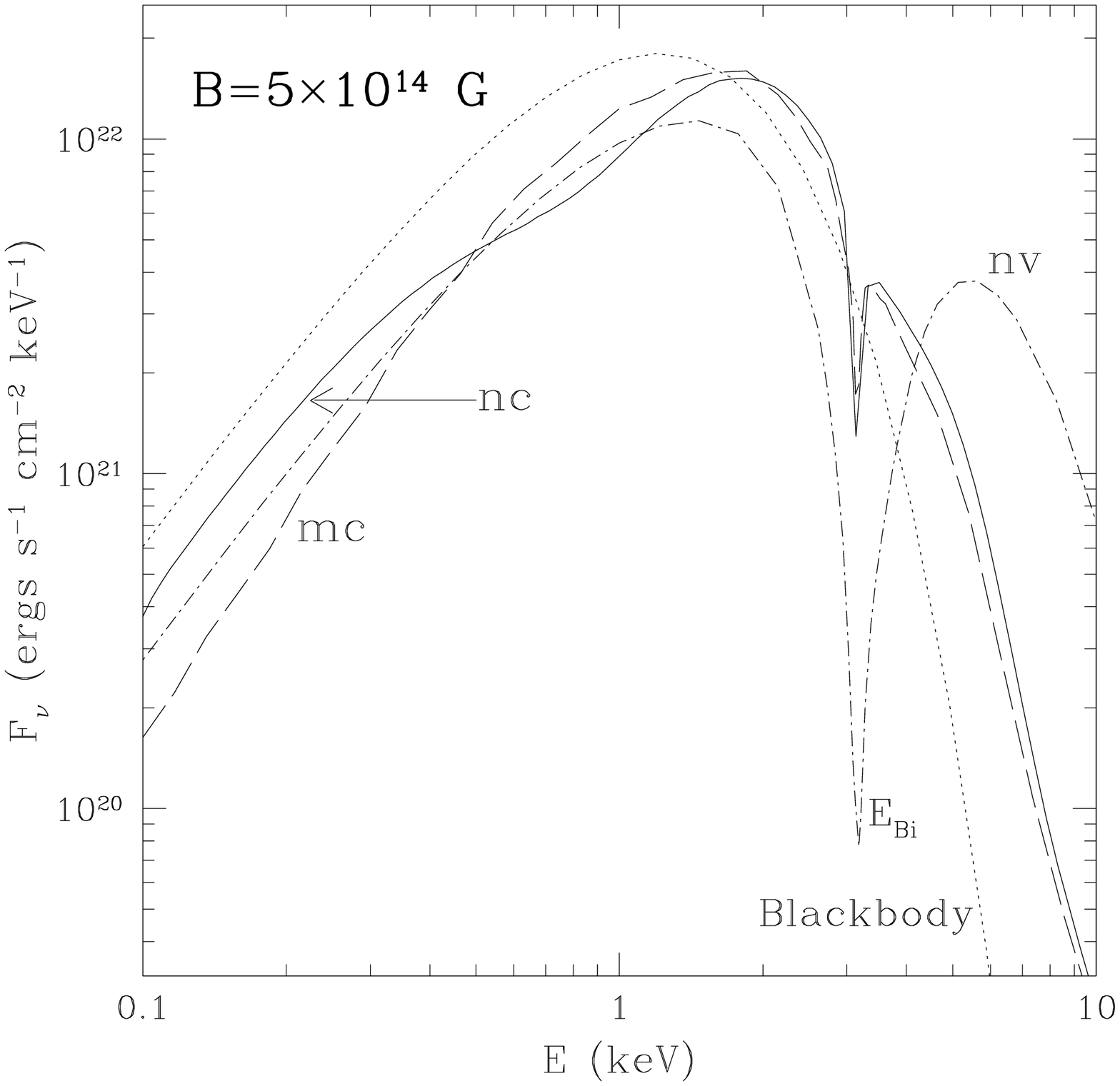}
\caption{
Spectra of fully ionized hydrogen atmospheres with
$B=5\times 10^{14}$~G, $\Thetab=0$, and $\Teff=5\times 10^6$~K.
The solid line is for an atmosphere with vacuum polarization
but no mode conversion (nc),
the dashed line is for an atmosphere with complete mode conversion (mc),
the dot-dashed line is for an atmosphere with no vacuum polarization (nv),
and the dotted line is for a blackbody with $T = 5\times 10^6$~K.
The $\Ebi=3.15$~keV ion cyclotron feature (from $\sim 2$ to $\sim 5$~keV)
has an equivalent width $\sim 2$~keV when vacuum polarization is not
included (model nv) but is only $\sim 0.1$~keV in the no mode conversion case
(model nc) and $\sim 0.09$~keV in the complete mode conversion case (model mc).
\label{fig:spectrum5h514}
}
\end{figure}

\begin{figure}
\centering
\includegraphics[height=8cm]{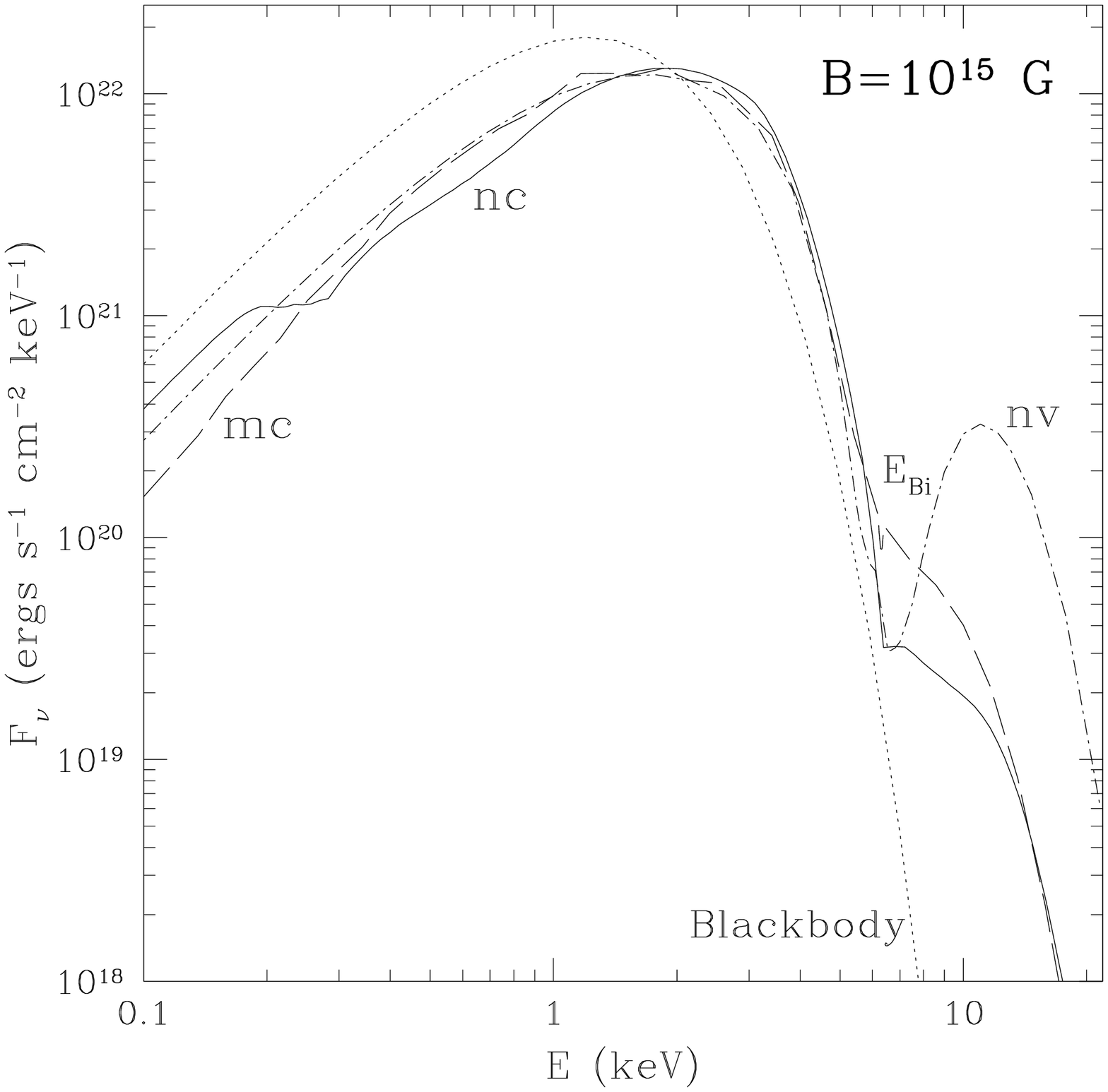}
\caption{
Spectra of fully ionized hydrogen atmospheres with
$B=10^{15}$~G, $\Thetab=0$, and $\Teff=5\times 10^6$~K.
The solid line is for an atmosphere with vacuum polarization
but no mode conversion (nc),
the dashed line is for an atmosphere with complete mode conversion (mc),
the dot-dashed line is for an atmosphere with no vacuum polarization (nv),
and the dotted line is for a blackbody with $T = 5\times 10^6$~K.
The $\Ebi=6.3$~keV ion cyclotron feature (from $\sim 5$ to $\sim 10$~keV)
has an equivalent width $\sim 4$~keV when vacuum polarization is not
included (model nv) but is only $\sim 0.2$~keV in the no mode conversion case
(model nc) and $\sim 0.03$~keV in the complete mode conversion case (model mc).
\label{fig:spectrum5h15}
}
\end{figure}

Figures~\ref{fig:spectrum5h14}-\ref{fig:spectrum5h15} depict
the spectra for fully ionized hydrogen atmosphere models with
$B=10^{14}$~G, $5\times 10^{14}$~G, and $10^{15}$~G, respectively,
$\Thetab=0$, and $\Teff=5\times 10^6$~K.
In each figure, the spectra from three models are depicted
together with a blackbody spectrum at $T=5\times 10^6$~K:
a model which neglects vacuum polarization (nv), a model which
includes vacuum polarization but no mode conversion (nc),
and a model which assumes complete vacuum polarization-induced
mode conversion (mc).
Recall that the nc model and the mc model represent two limiting
cases, and the true results with partial mode conversion
(see Section~\ref{sec:modeconv}) are expected to lie between the
nc and mc curves.  We see from
Figs.~\ref{fig:spectrum5h14}-\ref{fig:spectrum5h15} that the
differences in the spectra between the two limiting cases
are not significant.

Figures~\ref{fig:spectrum5h14}-\ref{fig:spectrum5h15} show that,
when vacuum polarization is neglected (nv), the spectra exhibit
significantly harder high energy tails and
a depletion of low energy photons relative to the blackbody.
The hard tails were already noted in previous studies of NS atmospheres
with $B\la 10^{13}$~G (e.g., Shibanov \etal~1992; Pavlov \etal~1995;
see also references cited in \holai),
and they arise because high energy photons have lower opacities
and thus decouple from deeper, hotter layers.
We see from Figs.~\ref{fig:spectrum5h14}-\ref{fig:spectrum5h15}
that vacuum polarization can significantly reduce these high
energy tails ($E\ga$~a few keV) and causes the spectra to be
softer compared to the models without vacuum polarization.
This reduction of the high energy tail is due to
the broad depression caused by the density-dependent
vacuum resonance feature, as discussed in Section~\ref{sec:densd}
(see Fig.~\ref{fig:densd}).

Figures~\ref{fig:spectrum5h14}-\ref{fig:spectrum5h15} also reveal
that vacuum polarization can significantly suppress the ion
cyclotron line feature in the spectra.  When vacuum polarization
is neglected, the ion cyclotron line is broad
(see \holai\ for more discussion of the feature; see also Zane \etal~2001),
and the equivalent widths are 0.14~keV ($\Ebi=0.63$~keV),
2~keV ($\Ebi=3.15$~keV), and 4~keV ($\Ebi=6.3$~keV) for
$B_{14}=1$, $5$, and $10$, respectively.
When vacuum polarization is included but mode conversion is neglected
(model nc), the equivalent widths become 20~eV ($\Ebi=0.63$~keV),
0.1~keV ($\Ebi=3.15$~keV), and 0.2~keV ($\Ebi=6.3$~keV),
i.e., the equivalent widths of the ion cyclotron line have been
reduced by a factor of $\sim 10$;
when mode conversion is assumed to be complete (model mc), the
equivalent widths are even smaller:
6~eV ($\Ebi=0.63$~keV), 0.09~keV ($\Ebi=3.15$~keV), and 0.03~keV
($\Ebi=6.3$~keV),
i.e., a reduction by a factor $> 10$.
This reduction of the ion cyclotron line strength by vacuum polarization
was expected from the analysis in Section~\ref{sec:densd}
(see Fig.~\ref{fig:densd}), and it occurs when the ion cyclotron
energy $\Ebi$ overlaps with the broad depression caused by the
vacuum resonance.  The reduced width of the ion cyclotron line
makes the line difficult to observe with current X-ray detectors
(see Section~\ref{sec:discussion}).

\begin{figure}
\centering
\includegraphics[height=8cm]{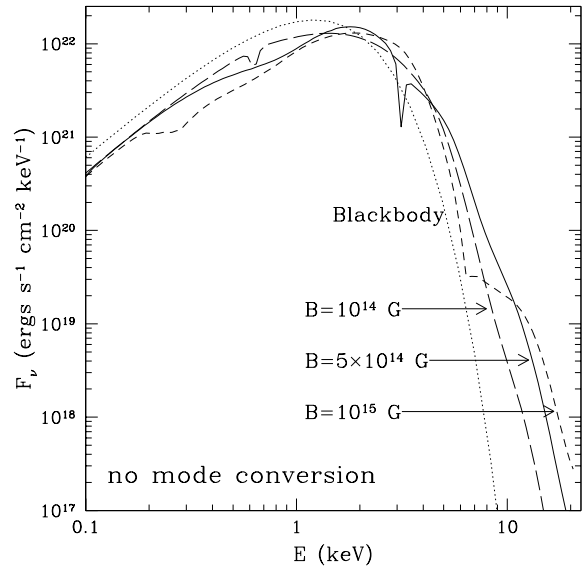}
\caption{
Spectra of fully ionized hydrogen atmospheres with
$\Thetab=0$ and $\Teff=5\times 10^6$~K when vacuum polarization is
included but mode conversion is neglected (nc).
The long-dashed line is for the $B=10^{14}$~G atmosphere,
the solid line is for the $B=5\times 10^{14}$~G atmosphere,
the short-dashed line is for the $B=10^{15}$~G atmosphere,
and the dotted line is for a blackbody with $T = 5\times 10^6$~K.
\label{fig:spectrum5hmagnc}
}
\end{figure}

\begin{figure}
\centering
\includegraphics[height=8cm]{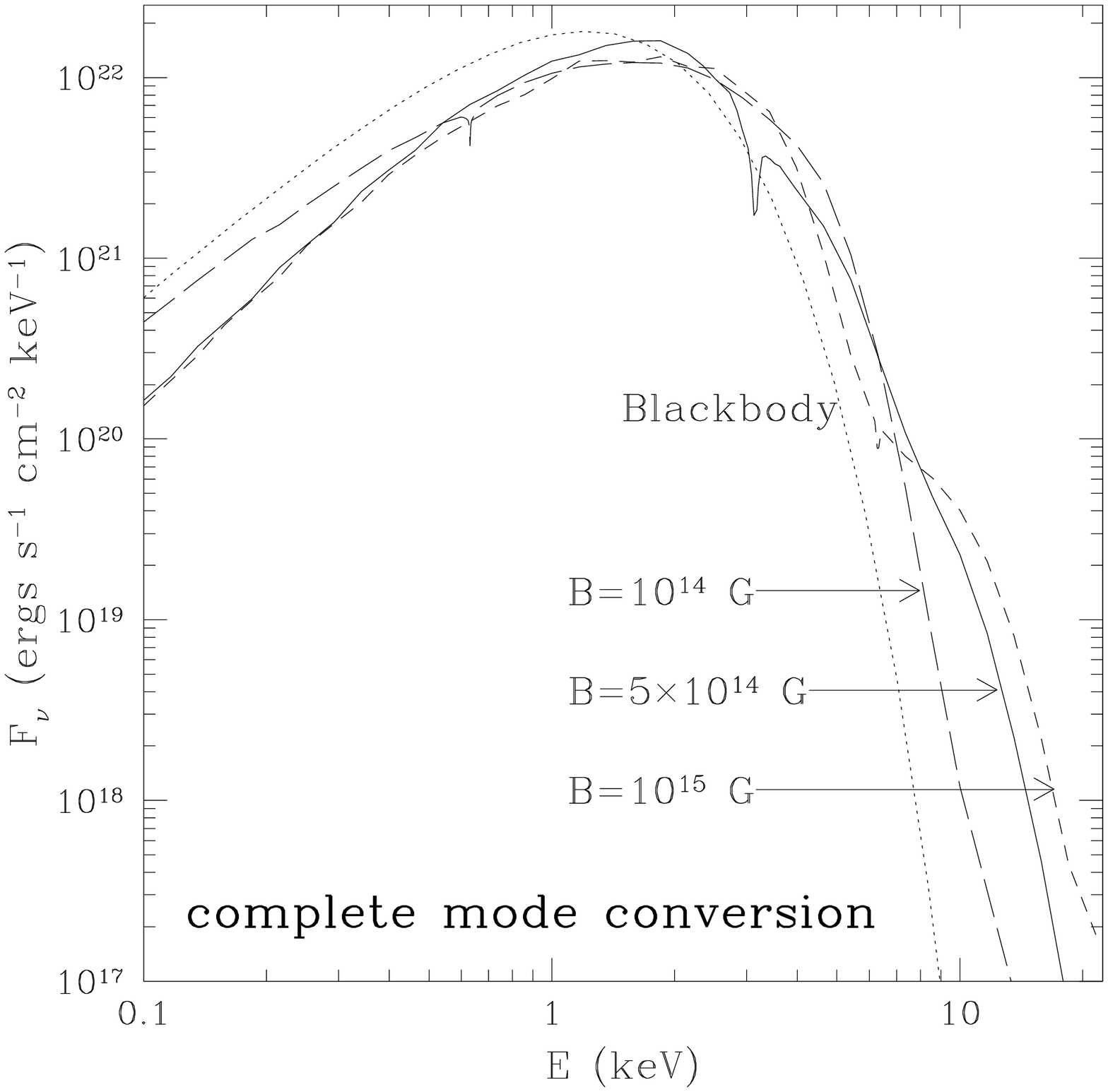}
\caption{
Spectra of fully ionized hydrogen atmospheres with
$\Thetab=0$ and $\Teff=5\times 10^6$~K when mode conversion is
complete (mc).
The long-dashed line is for the $B=10^{14}$~G atmosphere,
the solid line is for the $B=5\times 10^{14}$~G atmosphere,
the short-dashed line is for the $B=10^{15}$~G atmosphere,
and the dotted line is for a blackbody with $T = 5\times 10^6$~K.
\label{fig:spectrum5hmagmc}
}
\end{figure}

A comparison of the spectra for the models with different magnetic
fields is shown in Figure~\ref{fig:spectrum5hmagnc} for the case which
includes vacuum polarization but no mode conversion (nc) and in
Figure~\ref{fig:spectrum5hmagmc} for the case which includes
vacuum polarization with complete mode conversion (mc).
Note that despite the suppression of the high energy tails by the
vacuum polarization effect, all spectra are still harder than the
blackbody spectrum.

\subsection{Comparison with Previous Work} \label{sec:saturation}

As mentioned in Section~\ref{sec:intro}, there have been few previous
works on magnetar atmosphere models that include the effect of
vacuum polarization.  The most recent and relevant one is that
by \"{O}zel~(2001), who constructed self-consistent atmosphere
models with parameters similar to those described in our paper, i.e.,
$\Teff\sim 5\times 10^6$~K, $B\sim 10^{14}-10^{15}$~G, $\Thetab=0$,
and the atmosphere consists of fully ionized hydrogen.
While there are qualitative resemblances between some aspects of our
results and \"{O}zel's, e.g., the temperature profiles show plateau
features which are weakened by vacuum polarization
(compare our Fig.~\ref{fig:tempgrid} to her Fig.~5),
there are also major differences
in the physical ingredients and methods used in our models and
those of \"{O}zel's.
(1) \"{O}zel applied the vacuum polarization formulae which
are valid only for $B\la\Bq=4.4\times 10^{13}$~G.  At $B\ga 10^{15}$~G,
these incorrect formulae underestimate the vacuum resonance
energy by a factor of a few (see Fig.~\ref{fig:evp}).
(2) \"{O}zel neglected ions in the plasma response and the ion
cyclotron resonance in the opacities.  As discussed in \holai, the
ion effects cannot be neglected since they influence the spectrum
in the same energy range as vacuum polarization effects.
Figure~\ref{fig:spectrum5hnoionvp} compares the spectra of
atmosphere models which include ions and neglects ions.
Despite the suppression of the ion cyclotron feature by the
vacuum polarization effect (Section~\ref{sec:spectra}), the
``with ion'' and ``no ion'' cases have appreciably different
spectra (e.g., at $E=10$~keV, the flux of model ``nc'' differs
from model ``nc/no ions'' by a factor of about ten).
(3) \"{O}zel's work neglects the mode
conversion effect due to vacuum polarization
(\laiho; Section~\ref{sec:modeconv}).
(4) Even when mode conversion is neglected, \"{O}zel's method of
treating the vacuum resonance differs from ours.  As discussed
in Section~\ref{sec:grid}, to fully account for the density-dependent
vacuum resonance feature in the opacity (and thus producing smooth
spectra), one would need to have a prohibitively high density/depth
grid resolution.
\"{O}zel produces smooth spectra by adopting a ``saturation'' method,
whereby the peak X-mode opacity is restricted by requiring
$|\polarbvp|\ge 0.01$
[see eqs.~(\ref{eq:polarbvp0}) and (\ref{eq:vpbetares}); recall
that the vacuum resonance occurs at $\polarbvp=0$].
However, as discussed in Section~\ref{sec:opacity}, the region
with $|\polarbvp|\sim |E/\Evp-1|\la x_{\rm V}\sim\uel^{-1/2}$
[$\sim 0.001$ at 1~keV for $B=10^{14}$~G; see eq.~(\ref{eq:vpwidth1})
and Fig.~\ref{fig:vpline}]
contributes most to the ``equivalent width'' of the vacuum
resonance feature.  Therefore the saturation method does not
capture the main contribution of the vacuum resonance feature in
the opacities and significantly underestimates the effect of
vacuum polarization.
Figure~\ref{fig:tempgrid} shows the temperature profile of the
atmosphere model using our ``equal grid'' method together with
the restriction $|\polarbvp|\ge 0.01$.  Clearly, since the
saturation scheme does not capture the entire effect of the
opacity change, the heating of the atmosphere (compared to the
case in which vacuum polarization is neglected) due to the
vacuum resonance feature is less than the model with no
saturation restriction.
Figure~\ref{fig:spectrum5hsat} compares the spectra with and
without saturation.
Two effects are responsible for the qualitative differences.
First, because saturation underestimates the ``equivalent width''
of the vacuum resonance feature, it does not yield the true decrease
in flux (see Section~\ref{sec:densd}) so that the
the broad depression does not reach as low of energies.  Second,
the softer spectra of the saturation model at energies $E\ga 1$~keV
is the result of the lower temperature at the photon decoupling layers
in this model (see Fig.~\ref{fig:tempgrid}).
We also note again that even though vacuum polarization
produces harder spectra at high energies than the blackbody spectrum at
the same effective temperature, the spectra are still softer than
atmosphere models which neglect vacuum polarization
(see Figs.~\ref{fig:spectrum5h14}-{\ref{fig:spectrum5h15}).

\begin{figure}
\centering
\includegraphics[height=8cm]{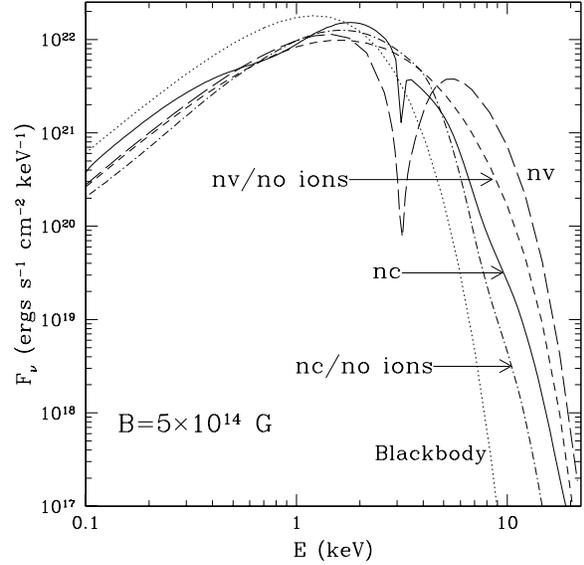}
\caption{
Spectra of fully ionized hydrogen atmospheres with
$B=5\times 10^{14}$~G, $\Thetab=0$, and $\Teff=5\times 10^6$~K.
The solid line is for an atmosphere with vacuum polarization
but no mode conversion (model nc, which includes ions),
the dot-dashed line is for an atmosphere with vacuum polarization
but no mode conversion and neglecting ion effects (nc/no ions),
the long-dashed line is for an atmosphere with no vacuum polarization
(model nv, which includes ions),
the short-dashed line is for an atmosphere with no vacuum polarization
and neglecting ion effects (nv/no ions),
and the dotted line is for a blackbody with $T = 5\times 10^6$~K.
\label{fig:spectrum5hnoionvp}
}
\end{figure}

\begin{figure}
\centering
\includegraphics[height=8cm]{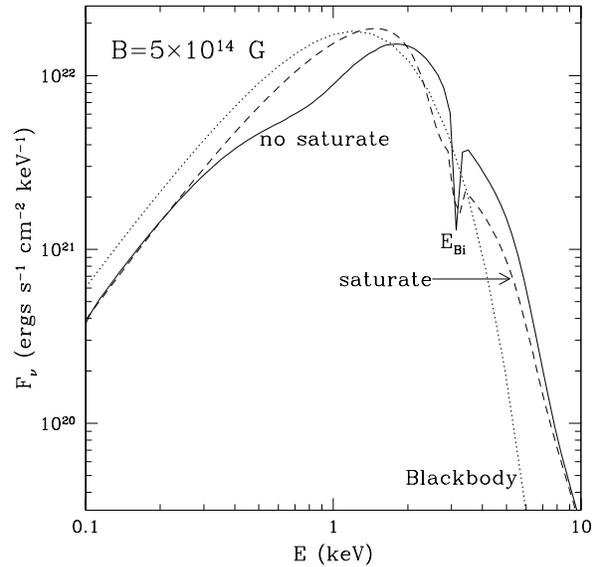}
\caption{
Spectra of fully ionized hydrogen atmospheres with
$B=5\times 10^{14}$~G, $\Thetab=0$, and $\Teff=5\times 10^6$~K.
The solid line (no saturate) is for an atmosphere with vacuum polarization
but no mode conversion using the ``equal grid'' method described
in Section~\ref{sec:grid},
the dashed line (saturate) is for the same atmosphere with vacuum polarization
but no mode conversion using the ``saturation'' method described
in Section~\ref{sec:saturation},
and the dotted line is for a blackbody with $T = 5\times 10^6$~K.
\label{fig:spectrum5hsat}
}
\end{figure}

It is appropriate to reiterate that our own numerical treatment
of the vacuum polarization effect also has limitations, and these
are discussed in Section~\ref{sec:numcomp}.
We also note that \"{O}zel's calculations include the full angular
dependence of the scattering opacity, whereas the results presented
in this paper are based upon an approximate calculation of the
scattering source function in order to speed up the computation
(see Section~2.2 of \holai).  Our numerical tests show that
this approximation does not affect the temperature profile of the
atmosphere and only produces a small reduction (by less than a
factor of two) in the flux at very high energies ($E\ga 10$~keV).
Finally, we have not examined the angular dependence and beaming
pattern of the emission as done in \"{O}zel~(2001)
(see also \"{O}zel, Psaltis, \& Kaspi~2001 and \"{O}zel~2002 for
analyses of the resulting pulse profiles and pulse fractions).

\setcounter{equation}{0}
\section{Discussion and Conclusion}
\label{sec:discussion}

We have presented a detailed study on the atmospheres
and thermal radiation of isolated neutron stars with superstrong
magnetic fields $B\ge 10^{14}$~G and effective temperatures
$\Teff\sim \mbox{a few}\times 10^6$~K.
Such a study is needed since surface emission has already been
detected from a number of magnetar candidates (AXPs and SGRs),
and current/future observations, when combined with theoretical modeling,
can potentially provide important constraints on the properties of these
enigmatic objects (see Section~\ref{sec:intro}).

Following up our previous work
on magnetized neutron star atmospheres (\holai), we focus on
the effect of vacuum polarization in this paper.  It was already
known from earlier theoretical studies (see Section~\ref{sec:intro}
and M\'{e}sz\'{a}ros~1992 for references), which we have generalized
to the $B\ga 10^{14}$~G regime and included the effect of ions,
that vacuum polarization changes the dielectric property of the plasma
and gives rise to a resonance feature in the opacity at photon energy
$\Evp\approx 1.02\,\Ye^{1/2}\rho_1^{1/2}B_{14}^{-1}f(B)$~keV, where
$\Ye=Z/A$ is the electron fraction,
$\rho_1=\rho/(\mbox{1 g cm$^{-3}$})$, $B_{14}=B/(\mbox{10$^{14}$ G})$,
and  $f(B)$ is a slowly-varying function of $B$ of order unity
(see Fig.~\ref{fig:evp}).
Furthermore, it was shown in \laiho\ (see also Gnedin \etal~1978;
Pavlov \& Gnedin~1984) that photons
with energies $E\ga$ a few keV propagating in the atmospheric plasma
can adiabatically convert from one polarization mode into another
at the vacuum resonance density
$\densvp\approx 0.96\,\Ye^{-1}E_1^2B_{14}^2f(B)^{-2}$~g~cm$^{-3}$,
where $E_1=E/(\mbox{1 keV})$;
this resonant mode conversion greatly influences the radiative transfer
because the two modes have vastly different opacities in strong magnetic
fields. In this paper, we have attempted to incorporate these vacuum
polarization effects into self-consistent atmosphere models.
Through both analytic considerations (Section~\ref{sec:densd};
see also \laiho) and numerical calculations (Section~\ref{sec:results}),
we have shown that vacuum polarization leads to
a broad depression in the high-energy ($E\ga$ a few keV)
radiation flux from the atmospheres as compared to models
without vacuum polarization.  Despite the rather sharp
vacuum resonance feature in the opacity at a given density,
the depression in the spectrum is broad (from a few keV to tens of keV,
depending on the field strength) because of the large density gradient
in the atmosphere.
As a result, the atmosphere emissions possess
softer high energy tails than models without vacuum polarization\footnote{
For all the models studied here, the spectra are still harder than the
blackbody spectrum (with the same effective temperature) because of
the non-grey opacities.  Also note that in this paper we are only
concerned with thermal emission from the neutron star surface.
Nonthermal emission or the reprocessing of the thermal emission
by the magnetospheric plasma may give rise to high energy, power-law
tails in the magnetar spectra (e.g., Thompson, Lyutikov, \& Kulkarni~2002).}.

Another important effect of vacuum polarization on the atmospheric
spectra is that the strength of the ion cyclotron line is greatly
suppressed when vacuum polarization is included in the atmosphere models
(see Figs.~\ref{fig:spectrum5h14}-\ref{fig:spectrum5h15}).
This is a direct consequence of the aforementioned flux
depression caused by vacuum polarization (see Fig.~\ref{fig:densd}):
when the depression trough overlaps with the ion cyclotron line at
$\Ebi=0.63\,\Ye B_{14}$~keV, i.e., for
$10^{14}\mbox{ G}\la B \la 5\times 10^{15}\mbox{ G}$,
the continuum flux around $\Ebi$ is greatly reduced, and
the ion cyclotron line appears narrower (see Section~\ref{sec:densd}).
For example, we find that the equivalent width of the $3.15$~keV proton
cyclotron line for $B=5\times 10^{14}$~G is reduced from
about 2~keV (no vacuum polarization) to about 0.1~keV
(with vacuum polarization).
Obviously the reduction of the ion cyclotron line width
makes the line more difficult to detect. Indeed, recent observations of
several AXPs with {\it Chandra} and {\it XMM-Newton} X-ray
telescopes failed to resolve any significant line features in the spectra
(e.g., Patel \etal~2001; Juett \etal~2002; Tiengo \etal~2002)\footnote{
For AXP 4U0142+61, Juett \etal~(2002) give an upper limit
of about 10-50~eV for the equivalent width of any line feature in the spectrum;
this is comparable to our predicted value of $\sim 6-20$~eV
at $B=10^{14}$~G and $\sim 100$~eV
at $B=5\times 10^{14}$~G.  In addition, it should be noted that the spectra
presented in our paper correspond to emission from a local patch of the
neutron star surface.  When the spectra are integrated over
the entire observable surface, which necessarily involves (very uncertain)
variations in the magnetic field strength and direction and
effective temperature, the equivalent width of the cyclotron line will be
further reduced.}; this could possibly be indicating the presence of vacuum
polarization effects.

We note that the radiative transfer formalism adopted in this paper
(and in previous atmosphere models by other researchers) relies on
the transfer of two photon polarization modes. This is inadequate
for treating the vacuum polarization-induced resonant mode conversion effect,
especially because the effectiveness of mode conversion depends on
photon energy (see Section~\ref{sec:modeconv} and \laiho).
The atmosphere models studied in this paper represent two
limiting solutions to the transfer
problem (no mode conversion and complete mode conversion), and they are
expected to bracket the true solution. Nevertheless, to properly
account for the mode conversion effect associated with vacuum polarization,
as well as mode collapse and the breakdown of Faraday depolarization
(see Section~\ref{sec:mcp}),
one must go beyond the modal description of the radiation field by formulating
and solving the transfer equation in terms of the photon density
matrix (or Stokes parameters) and including the effect of a nontrivial
refractive index.  We plan to study this problem in the future
(see also Lai \& Ho~2002b).

We caution that our models, as well as previous models of magnetar
atmospheres (see Section~\ref{sec:intro}), assume that the atmospheres
are completely ionized.  For the magnetic field strengths and
surface temperatures
characteristic of AXPs and SGRs, it is not clear that
bound atoms and molecules have sufficiently small
abundances to contribute negligibly to the opacity
(see the discussion section in \holai), given their greatly
enhanced binding energies in strong magnetic fields
(see Lai~2001 for a review).
However, it may be the case that, like the ion cyclotron line,
absorption lines due to bound species are suppressed if they fall
within the broad flux depression caused by vacuum polarization.
The absence of any resolved features in recent observations of
several AXPs (Patel \etal~2001; Juett \etal~2002; Tiengo \etal~2002)
may therefore be indicative of the vacuum polarization effect in
superstrong magnetic fields.
Also of concern is the
dense plasma effect on the radiative transfer (see \holai).
Clearly, much work remains to be done before we can have complete
confidence in our quantitative understanding of the radiation
from the surfaces of strongly magnetized neutron stars.

\section*{Acknowledgments}

We thank Tomasz Bulik, Lars Hernquist, Jeremy Heyl,
Alex Potekhin, Ira Wasserman, Silvia Zane, and especially George
Pavlov for useful discussion and correspondence.
We thank Roberto Turolla for useful comments and questions, which
led us to expand a footnote in the earlier version to the new
Section~\ref{sec:mcp}.
We are grateful to the Cornell Hewitt Computer Laboratory for the
use of their facilities.
This work is supported in part by NASA grant
NAG 5-8484 and NSF grant AST 9986740.
W.C.G.H. is also supported by a fellowship from the
NASA/New York Space Grant Consortium, and D.L. is also supported
by a fellowship from the A.P. Sloan Foundation.


\appendix

\setcounter{equation}{0}
\section{Atmosphere Including Vacuum Polarization Resonance Effects:
Toy Models}
\label{sec:toy}

\subsection{Opacities}
\label{sec:toyopacity}

In Section~\ref{sec:densd}, we examined the effect of the vacuum
polarization resonance feature on the depth at which the photons
that comprise the surface spectrum are emitted.  To determine the
emission spectrum, the temperature profile must be determined
self-consistently.  In this section, we consider several toy atmosphere
models based on the simplified opacities shown in
Figure~\ref{fig:opacitytoy}.  These models serve to illustrate
the key physical effects of the vacuum resonance feature and
the numerical subtleties when constructing real atmosphere models
(Sections~\ref{sec:numcomp} and \ref{sec:results}).

\begin{figure}
\centering
\includegraphics[height=8cm]{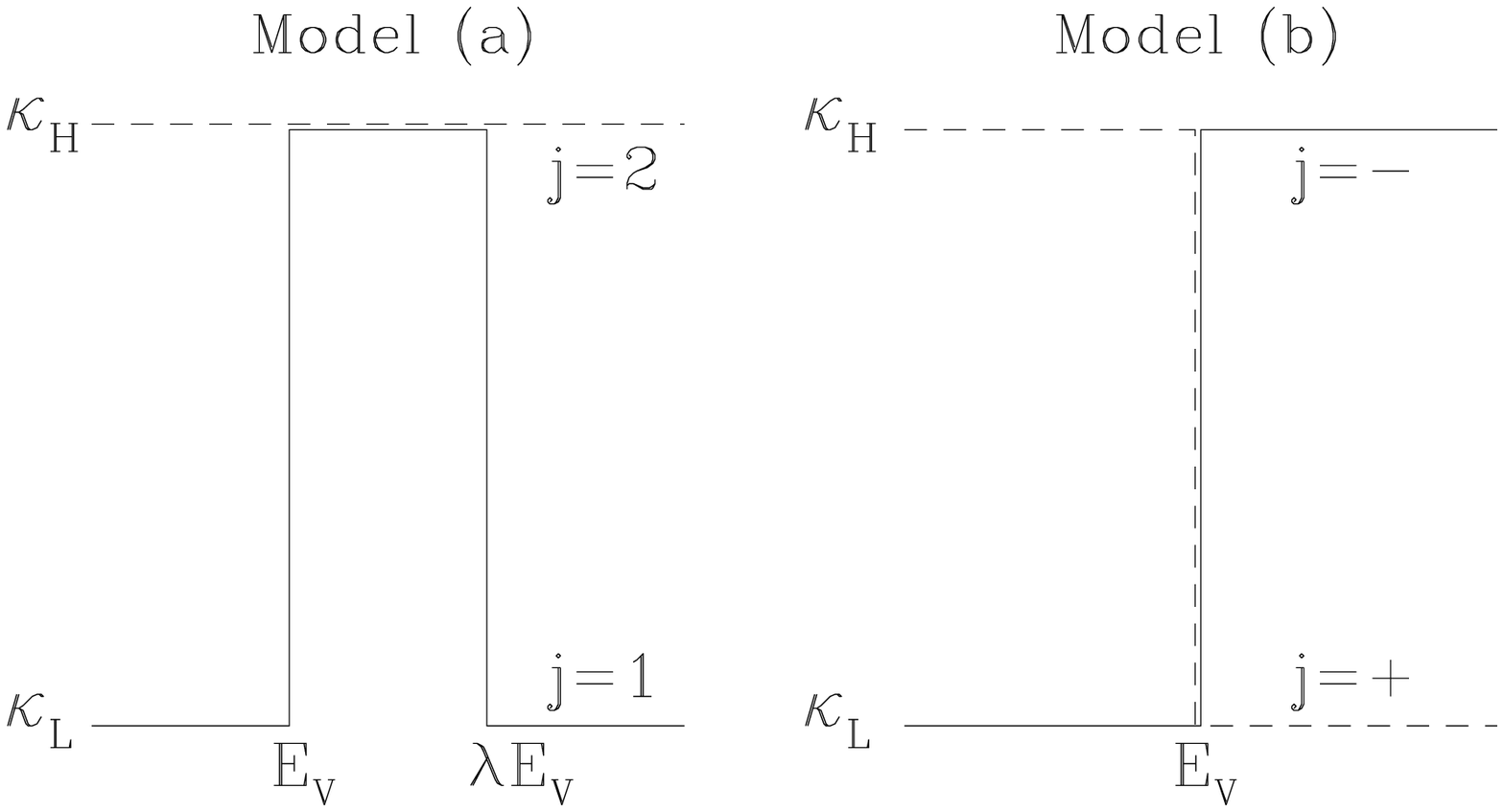}
\caption{
Toy model opacities used in Appendix~\ref{sec:toy} with
$j=1,2$ indicating the X and O-mode, respectively,
and $j=\pm$ indicating the plus and minus-modes, respectively.
Model~(a) mimics the case where mode conversion is neglected
[see eq.~(\ref{eq:toyopacityen})];
Model~(b) mimics the case where mode conversion is complete
[see eq.~(\ref{eq:toyopacityenmc})].
The opacities are parameterized by the location of the resonance
feature $\Evp$ [eq.(\ref{eq:evptoy})],
the width parameter $\lambda$, and $\kaph$ and $\kapl$.
\label{fig:opacitytoy}
}
\end{figure}

As shown in Section~\ref{sec:opacity}, in a real atmosphere,
the O-mode ($j=2$) opacity is largely unaffected by the vacuum
resonance while the X-mode ($j=1$) opacity exhibits a significant
peak around $\Evp$.
Therefore, for Model~(a), we assume the O-mode opacity is grey,
$\kappaE^O=\kaph$, and the X-mode opacity has a simple
square function form:
\be
\kappaE^X = \left\{ \begin{array}{ll}
 \kapl & \mbox{for } E<\Evp \mbox{ or } E>\lambda\Evp \\
 \kaph & \mbox{for } \Evp\le E\le \lambda\Evp \end{array} \right.
 \, \mbox{[Model (a)]} ,
 \label{eq:toyopacityen}
\ee
where
\ba
\Evp= \zeta\rho^{1/2}, \label{eq:evptoy}
\ea
$\rho$ is in g cm$^{-3}$, $E$ is in keV,
$\zeta\approx B_{14}^{-1}f(B)$ [see eqs.~(\ref{eq:evp0})-(\ref{eq:evplohi})],
and the width of the line feature is $(\lambda-1)\Evp$.
Equation~(\ref{eq:toyopacityen}) thus resembles the X-mode opacity
shown in the left panels of Fig.~\ref{fig:ken} near the vacuum
polarization resonance.
From Section~\ref{sec:opacity}, we know that $\kaph/\kapl\sim\uel$ at
the resonance peak, and the width is
$(\lambda-1)\sim (\uel^{1/2}|1-\uion|)^{-1}$; we will see that
only the ``equivalent width''
$(\kaph/\kapl)(1-\lambda^{-2})\sim\uel^{1/2}/|1-\uion|$
($\sim$$10^3$ at 1~keV for $B=10^{14}$~G) matters for the radiation
spectrum.
Model~(a) mimics the real atmosphere models where vacuum-induced
mode conversion is neglected.

In Model~(b), we assume the plus and minus-mode opacities are simple
step functions:
\ba
\kappaE^+ & = & \left\{ \begin{array}{ll}
 \kaph & \mbox{for } E<\Evp \\
 \kapl & \mbox{for } E \ge \Evp \end{array} \right.
 \qquad \mbox{[Model (b)]} \nonumber \\
\kappaE^- & = & \left\{ \begin{array}{ll}
 \kapl & \mbox{for } E<\Evp \\
 \kaph & \mbox{for } E \ge \Evp \end{array} \right.
 \qquad \mbox{[Model (b)]} .
 \label{eq:toyopacityenmc}
\ea
These opacities resemble the behavior of the plus and minus-modes in
the right panels of Fig.~\ref{fig:ken}.  Thus Model~(b) mimics
the atmospheres where mode conversion is assumed to be complete.

\begin{figure}
\centering
\includegraphics[height=8cm]{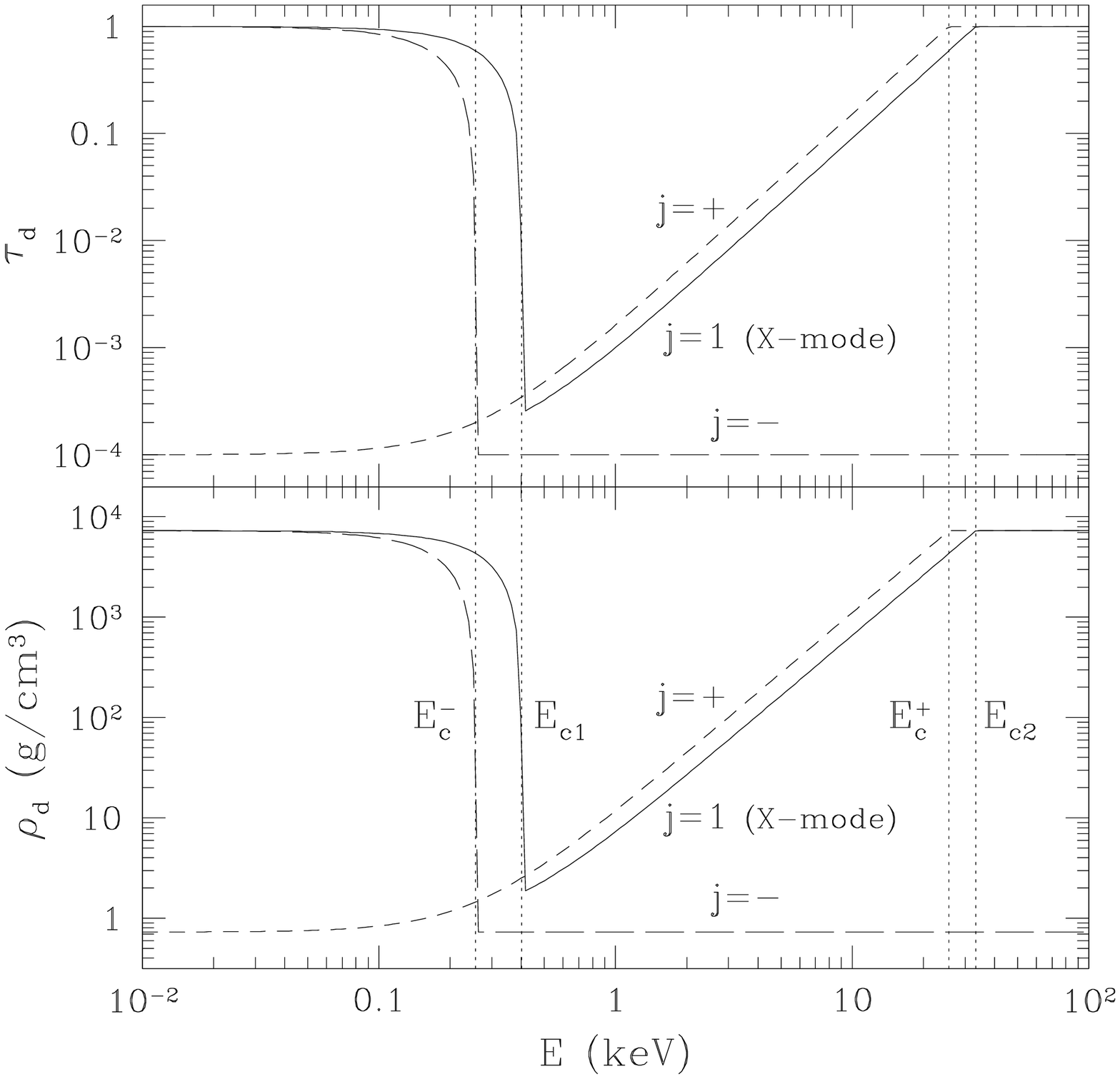}
\caption{
The upper panel shows the optical depth $\taud$ (in terms of
the opacity $\kapl$) at which photons of different modes
decouple from the matter as a function of energy for $T=5\times 10^6$~K,
$\zeta=0.3$, $\lambda=1.3$, $\kaph=\keso$, and $\kapl=10^{-4}\keso$.
The lower panel shows the corresponding density $\rhod$ at
these depths.
The solid lines are for the X-mode with an opacity given by Model~(a)
(the O-mode decoupling depth $\taud=1$ is not shown),
and the short and long-dashed lines are for Model~(b)
(see Fig.~\ref{fig:opacitytoy}).
The vertical dotted lines indicate the critical energies $E_{\rm c1}$
[eq.~(\ref{eq:toyec1})], $E_{\rm c2}$ [eq.~(\ref{eq:toyec2})],
$E^+_{\rm c}$ [eq.~(\ref{eq:toyecp})],
and $E^-_{\rm c}$ [eq.~(\ref{eq:toyecm})].
\label{fig:toytaud}
}
\end{figure}

\subsection{Photon-Matter Decoupling Region}
\label{sec:toydensd}

As in Section~\ref{sec:densd}, we can calculate the density $\rhod$
where photons of different modes decouple from the matter.
The corresponding decoupling depth $\taud(E)$ is determined
from $1 = \int_0^{\taud}\!d\tauo[\kappaE(\tauo)/\kapl]$,
where $d\tauo=-\rho\kapl\,dz$.
Similar to Section~\ref{sec:densd}, we assume a constant
temperature profile when calculating $\taud$ and $\rhod$
(see Section~\ref{sec:densd} for justification).
Thus, for a fully ionized hydrogen atmosphere,
$\rho(\tauo)=1.45\,yT_6^{-1}=\rho_0\tauo T_6^{-1}$ ~g~cm$^{-3}$,
where $\rho_0\approx 1.45/\kapl$ (here $\kapl$ is in cm$^2$g$^{-1}$ and
$\rho_0$ is in g~cm$^{-3}$).
Combining the density profile with equation~(\ref{eq:toyopacityen})
gives the X-mode opacity as a function of $\tauo$, i.e.,
$\kappaE(\tauo)=\kaph$ for
$(T_6/\rho_0)(E/\lambda\zeta)^2\le\tauo\le(T_6/\rho_0)(E/\zeta)^2$
and $\kappaE(\tauo)=\kapl$ otherwise.
We then obtain the decoupling depth $\taud$ of the X-mode
as a function of energy,
\be
\taud(E) = \left\{ \begin{array}{l}
 1 - \lp\kaph/\kapl-1\rp\lp 1-\lambda^{-2}\rp
 \rho_0^{-1}T_6\lp E/\zeta\rp^2 \\
 \qquad \qquad \qquad \mbox{for } E<E_{\rm c1} \\
 \kapl/\kaph + \lp 1-\kapl/\kaph\rp
 \rho_0^{-1}T_6\lb E/\lp\zeta\lambda\rp\rb^2 \\
 \qquad \qquad \qquad \mbox{for } E_{\rm c1}\le E\le E_{\rm c2} \\
 1 \qquad \mbox{ for } E>E_{\rm c2}
 \end{array} \right.. \label{eq:toytaud2}
\ee
The two critical energies, $E_{\rm c1}$ and $E_{\rm c2}$,
corresponding to when the decoupling depth $\taud=(T_6/\rho_0)(E/\zeta)^2$
and $\taud=(T_6/\rho_0)(E/\lambda\zeta)^2$ are
\ba
E_{\rm c1} & = & \zeta\rho_0^{1/2}T_6^{-1/2}\lambda \lb 1
 + \frac{\kaph}{\kapl} \lp\lambda^2-1\rp\rb^{-1/2} \label{eq:toyec1} \\
E_{\rm c2} & = & \zeta\rho_0^{1/2}T_6^{-1/2}\lambda. \label{eq:toyec2}
\ea
The decoupling density, $\rhod=\rho(\taud)$, is given by
\be
\rhod(E) = \left\{ \begin{array}{l}
 \rho_0 T_6^{-1} - \lp\kaph/\kapl-1\rp\lp 1-\lambda^{-2}\rp
 \lp E/\zeta\rp^2 \\ \qquad \qquad \qquad \mbox{for } E<E_{\rm c1} \\
 \rho_0 T_6^{-1}\kapl/\kaph + \lp 1
 - \kapl/\kaph\rp\lb E/\lp\zeta\lambda\rp\rb^2 \\
  \qquad \qquad \qquad \mbox{for } E_{\rm c1}\le E\le E_{\rm c2} \\
 \rho_0 T_6^{-1} \qquad \mbox{for } E>E_{\rm c2} \end{array} \right..
 \label{eq:densdtoy}
\ee
Note that when $\kapl/\kaph\ll 1$ and $\lambda\approx 1$, the
decoupling density $\rhod(E)\approx(E/\zeta)^2=\densvp(E)$ for
$E_{\rm c1}\le E\le E_{\rm c2}$, which
is in agreement with the result of Section~\ref{sec:densd}
(see Fig.~\ref{fig:densd}).
Equations~(\ref{eq:toytaud2}) and (\ref{eq:densdtoy}) are plotted
in Figure~\ref{fig:toytaud} with $\zeta=0.3$, $\lambda=1.3$,
$\kaph=\keso=0.4$~cm$^2$~g$^{-1}$, and $\kapl=10^{-4}\keso$;
the choice of these parameter
values is discussed in Section~\ref{sec:numcomptoy}.
The minimum decoupling depth and density occur at $E_{\rm c1}$:
\ba
\taud(E_{\rm c1}) & = & \frac{1}{\lambda^{-2}+\kaph/\kapl
 \lp 1-\lambda^{-2}\rp}  \label{eq:toytaudec1} \\
\rhod(E_{\rm c1}) & = & \frac{\rho_0 T_6^{-1}}{\lambda^{-2}
 + \kaph/\kapl \lp 1-\lambda^{-2}\rp}. \label{eq:toyrhodec1}
\ea
Clearly the minimum decoupling depth depends critically on the
equivalent width of the opacity feature, i.e.,
$(\kaph/\kapl)(1-\lambda^{-2})$,
which corresponds to $\Gammavp$ in equation~(\ref{eq:vpwidth2}).
In fact, in the limit of $\kapl/\kaph\ll 1$ and $\lambda\approx 1$,
the parameters $\kapl$, $\kaph$, and $\lambda$ enter
equations~(\ref{eq:toytaud2})-(\ref{eq:toyrhodec1}) only through
the combination $(\kaph/\kapl)(1-\lambda^{-2})$.
The larger the equivalent width, the shallower the decoupling depth and
hence the cooler the region where the observable photons are
produced and the deeper the depression that is produced in the
emission spectrum.
Also, equations~(\ref{eq:toyec1}) and (\ref{eq:toyec2}) show that,
by increasing the magnetic field (or decreasing $\zeta$), the broad
depression is shifted to lower energies.

Now consider Model~(b) with opacities given by
equation~(\ref{eq:toyopacityenmc}).  Using the same procedure
as above, we obtain the decoupling densities for the plus and
minus-mode:
\be
\rhod^+(E) = \left\{ \begin{array}{l}
 \rho_0 T_6^{-1}\kapl/\kaph + \lp 1-\kaph/\kapl\rp \lp E/\zeta\rp^2 \\
 \qquad \qquad \mbox{for } E<E_{\rm c}^+ \\
 \rho_0 T_6^{-1} \qquad \mbox{for } E>E_{\rm c}^+ \end{array}\right.
 \label{eq:densdptoy}
\ee
\be
\rhod^-(E) = \left\{ \begin{array}{l}
 \rho_0 T_6^{-1} - \lp\kaph/\kapl-1\rp \lp E/\zeta\rp^2 \\
 \qquad \qquad \qquad \mbox{for } E<E_{\rm c}^- \\
 \rho_0 T_6^{-1}\kapl/\kaph \qquad \mbox{for } E>E_{\rm c}^- \end{array}
 \right. , \label{eq:densdmtoy}
\ee
where
\ba
E_{\rm c}^+ & = & \zeta\rho_0^{1/2}T_6^{-1/2} \label{eq:toyecp} \\
E_{\rm c}^- & = & \zeta\rho_0^{1/2}T_6^{-1/2}(\kapl/\kaph)^{1/2}.
 \label{eq:toyecm}
\ea
Equations~(\ref{eq:densdptoy}) and (\ref{eq:densdmtoy}), along
with the corresponding decoupling depths for the plus and minus-modes,
are also plotted in Fig.~\ref{fig:toytaud}.

\subsection{Numerical Method} \label{sec:numcomptoy}

Using the opacities of Models~(a) and (b)
in our magnetic diffusion atmosphere code (for simplicity and
computation speed)
described in \holai, we obtain self-consistent atmosphere
models for a given choice of $\zeta$, $\lambda$, $\kaph$, and $\kapl$.
The results presented in this section are somewhat modified for
a full, angle-dependent solution of the radiative transfer equation,
but since here we are more interested in the qualitative effects of
vacuum polarization, the diffusion models are adequate for illustrative
purposes.
Temperature corrections and deviations from radiative equilibrium
and constant flux are $\la 1\%$ (see \holai).

For illustrative purposes, we choose the parameters $\zeta$, $\lambda$,
$\kaph$, and $\kapl$ for the toy models that roughly correspond to
the case of an atmosphere with magnetic field $B=5\times 10^{14}$~G.
For this magnetic field, equations~(\ref{eq:evp0})-(\ref{eq:evplohi})
give $\zeta\approx B_{14}^{-1}f(B)\approx 0.3$.
We choose a fairly broad width $(\lambda-1)=0.3$ for the feature
so that the required grid resolution is not too computationally
demanding.  We also choose values for the parameters $\kaph=\keso$
and $\kapl=10^{-4}\keso$ so that the equivalent width of the feature
$(\kaph/\kapl)(1-\lambda^{-2})=4000$, which corresponds roughly
to the equivalent width of the vacuum resonance feature for
$B=5\times 10^{14}$~G [see eq.~(\ref{eq:vpwidth3})].

As discussed in Section~\ref{sec:numcomp}, to correctly account
for the density-dependent opacity feature, such as in Model~(a),
in numerical calculations, it is important that the density or
depth grid is sufficiently dense so that the opacity features
at neighboring density grid points overlap.
For Model~(a) with $\lambda=1.3$, the minimum required grid
spacing is $|\Delta\rho|/\rho<(\lambda-1)^2=0.09$.  In our calculations,
we use 50 grid points per decade in $\tau$ (the total number of
depth grid points is then $D\approx 350$ covering seven decades
in $\tau$), which corresponds to $(\rho_{i}-\rho_{i-1})/\rho_i\sim 0.04-0.05$.
In addition, to obtain an accurate representation of the opacity
feature in the energy domain, we choose an energy grid that
is determined from the depth grid: for each depth grid point,
we place an energy grid point on either side of the opacity edge,
so that the total number of energy grid points $N=4D$ for Model~(a)
and $N=2D$ for Model~(b).  In this way, we completely characterize
the shape and strength of the opacity features in Models~(a) and (b).

\subsection{Numerical Results} \label{sec:toyresults}

Figure~\ref{fig:temptoy} plots the temperature profiles for
the atmosphere models with $\Teff=5\times 10^6$~K
using Model~(a) (mimicking the case of no mode conversion) and
Model~(b) (mimicking the case of complete mode conversion) and
$\zeta=0.3$, $\lambda=1.3$, $\kaph=\keso$, and $\kapl=10^{-4}\keso$.
Also shown is the temperature profile from a
double-grey atmosphere, i.e., two photon modes each with constant
opacities $\kappaE^X=\kapl$ for the X-mode and $\kappaE^O=\kaph$ for
the O-mode.
We see that both Model~(a) and Model~(b) have higher temperatures
(at all depths) than the double-grey case.
This is due to the heating effect associated with the increase in
opacity (see Section~\ref{sec:results} and Figs.~\ref{fig:tempgrid}
and \ref{fig:temp}).
Also note the temperature inversions that occur near $\tau\sim 600$;
this is due to the absorption of high-energy X-ray photons
as they encounter the resonance feature.
Figure~\ref{fig:spectrumtoy} shows the corresponding spectra of
the atmosphere models.
Note that all of these spectra are harder than a blackbody
at $T=5\times 10^6$~K and are similar to the behavior seen in
Fig.~\ref{fig:spectrum5h514}, except without the ion cyclotron line.

\begin{figure}
\centering
\includegraphics[height=8cm]{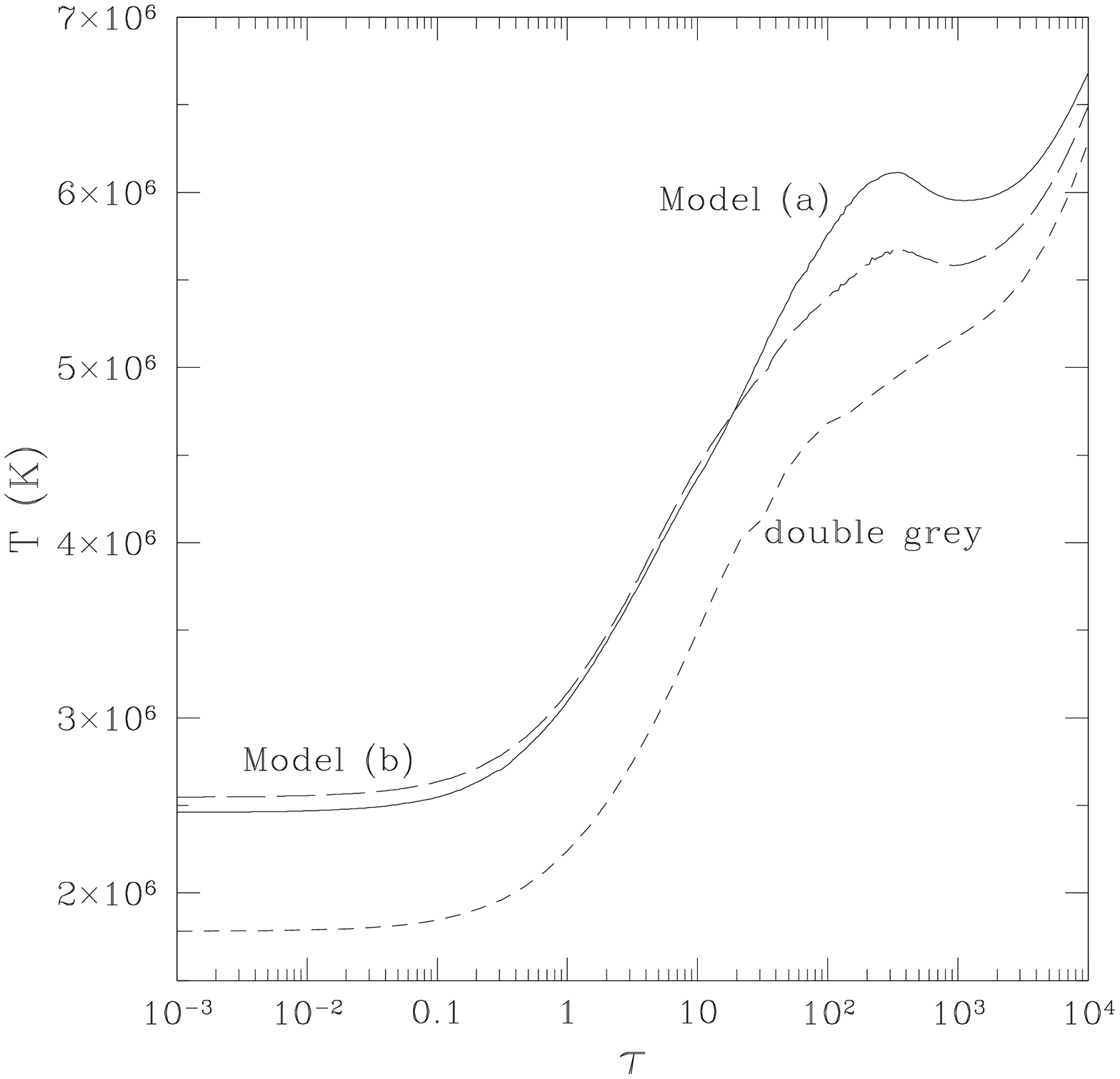}
\caption{
Temperature $T$ as a function of Thomson depth $\tau$
for atmosphere models using the toy opacities given in
Appendix~\ref{sec:toy} (see Fig.~\ref{fig:opacitytoy})
and $\Teff=5\times 10^6$~K,
$\zeta=0.3$, $\lambda=1.3$, $\kaph=\keso$, and $\kapl=10^{-4}\keso$.
The solid line is for the temperature profile of an atmosphere
with Model~(a), the long-dashed line is the profile for Model~(b),
and the short-dashed line is for a double grey atmosphere.
\label{fig:temptoy}
}
\end{figure}

\begin{figure}
\centering
\includegraphics[height=8cm]{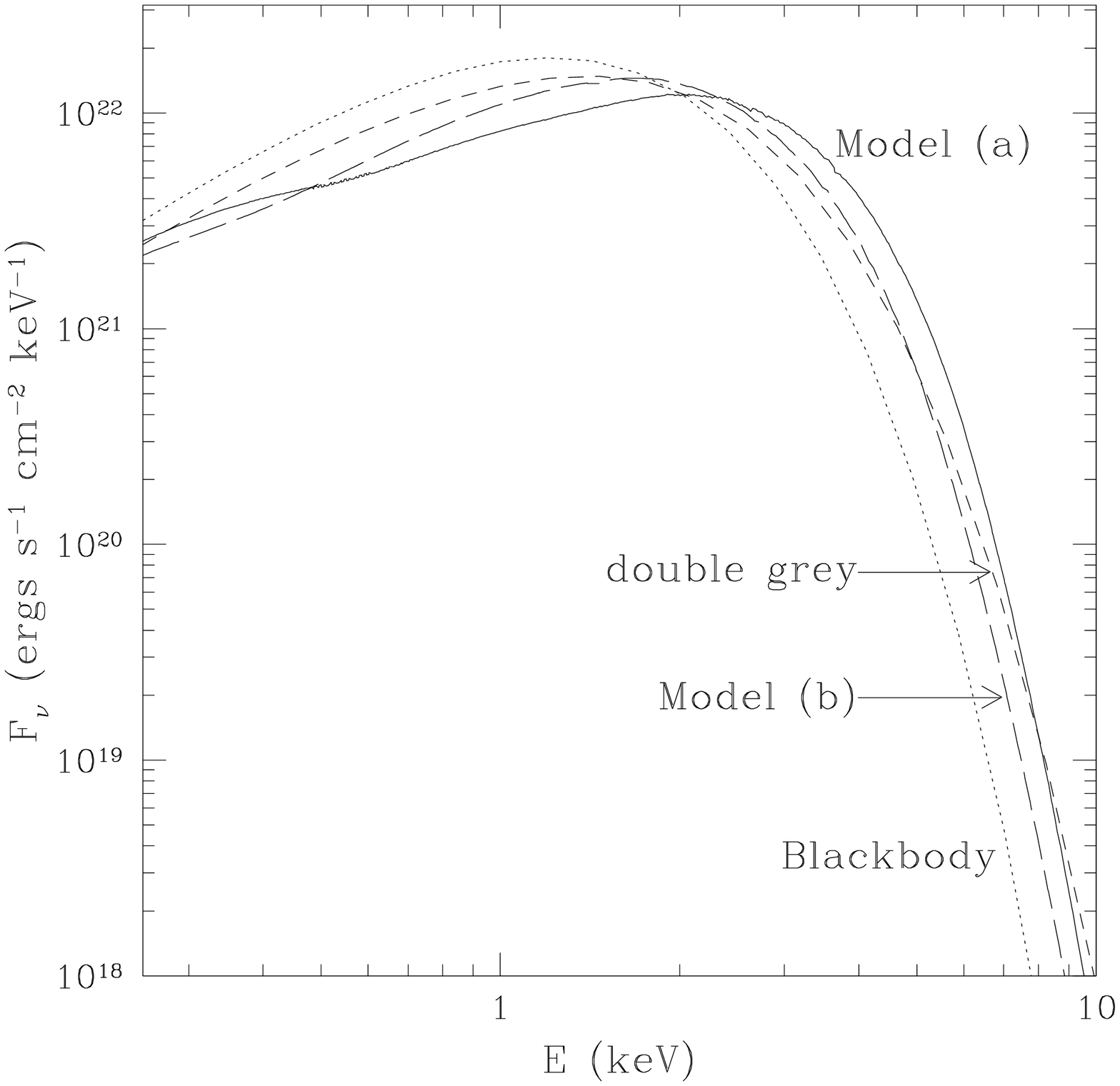}
\caption{
Spectra of atmospheres with $\Teff = 5\times 10^6$~K,
$\zeta=0.3$, $\lambda=1.3$, $\kaph=\keso$, and $\kapl=10^{-4}\keso$.
The solid line is for an atmosphere using the Model~(a) opacities,
the long-dashed line for an atmosphere using the Model~(b) opacities,
the short-dashed line is for a double grey atmosphere,
and the dotted line is for a blackbody with $T = 5\times 10^6$~K.
\label{fig:spectrumtoy}
}
\end{figure}

\label{lastpage}

\end{document}